\documentclass[a4paper,11pt]{article}
\usepackage{aaskaiid}
\usepackage{orcidlink} 
\usepackage[colorinlistoftodos]{todonotes}
\usepackage{bm}

\newcommand*\degr{\ensuremath{^\circ}}
\newcommand*\arcmin{\ensuremath{^\prime}}
\newcommand*\arcsec{\ensuremath{^{\prime\prime}}}

\newcommand*\farcs{\ensuremath{\overset{\prime\prime}{.}}}

\title{The magnetic field in the Milky Way Galaxy: from large to small scales}
\ShortTitle{Magnetic field in the Milky Way Galaxy}

\author[1]{Xiaohui Sun\,\orcidlink{0000-0002-3464-5128}}
\ShortName{Sun et al.} 
\author[2,10]{Jennifer West}
\author[3]{Marijke Haverkorn}
\author[4,5,6]{Andrea Bracco\,\orcidlink{0000-0003-0932-3140}}
\author[2]{Anna Ordog}
\author[7]{Sui Ann Mao}
\author[7]{Yik Ki Ma}
\author[1]{Wenhui Jing}
\author[8,9]{Thiem Hoang}

\affiliation[1]{School of Physics and Astronomy, Yunnan University, Kunming 650500, China}
\emailAdd{xhsun@ynu.edu.cn}
\affiliation[2]{Dominion Radio Astrophysical Observatory, Herzberg Astronomy \& Astrophysics, National Research Council Canada, P.O. Box 248, Penticton, BC V2A 6J9, Canada}

\emailAdd{jennifer.west@utas.edu.au}
\affiliation[3]{Department of Astrophysics/IMAPP, Radboud University, Nijmegen, PO Box 9010, 6500 GL Nijmegen, the Netherlands}
\emailAdd{m.haverkorn@astro.ru.nl}
\affiliation[4]{LUX, Observatoire de Paris, Université PSL, Sorbonne Université, CNRS, 75014 Paris, France}
\affiliation[5]{INAF–Osservatorio Astrofisico di Arcetri, Largo E. Fermi 5, 50125 Firenze, Italy}
\affiliation[6]{Laboratoire de Physique de l'Ecole Normale Sup\'erieure, ENS, Universit\'e PSL, CNRS, Sorbonne Universit\'e, Universit\'e de Paris, 75005 Paris, France}
\emailAdd{andrea.bracco@obspm.fr}
\emailAdd{anna.ordog@ubc.ca}
\affiliation[7]{ Max-Planck-Institut f\"ur Radioastronomie, Auf dem H\"ugel 69, D-53121 Bonn, Germany}
\emailAdd{mao@mpifr-bonn.mpg.de}
\emailAdd{ykma@mpifr-bonn.mpg.de}
\emailAdd{wh.jing@outlook.com}
\affiliation[8]{ Korea Astronomy and Space Science Institute, Daejeon 34055, Republic of Korea}
\affiliation[9]{Department of Astronomy and Space Science, University of Science and Technology, 217 Gajeong-ro, Yuseong-gu, Daejeon, 34113, Republic of Korea}
\emailAdd{thiemhoang@kasi.re.kr}
\affiliation[10]{
School of Natural Sciences, University of Tasmania, PO Box 807, Sandy Bay, TAS 7006, Australia}

\abstract{The Milky Way is the galaxy in which we can study its magnetic field to the finest details, providing an ideal laboratory to understand the fundamental questions: how magnetic field is generated and evolves, and how it influences other components in the Galaxy. An SKA-Mid polarization survey will produce an all-sky rotation measure (RM) grid with a density of about 100~deg$^{-2}$, which is approximately two orders of magnitude larger than what is currently available, and produce total intensity, polarized intensity, and RM all-sky images of diffuse emission covering scales from about $10\arcsec$ upward after combination with single-dish observations. The dense RM grid and images of diffuse emission will allow us to determine the most complete picture of the magnetic field in the southern Galactic hemisphere from large to small scales. }

\begin{document}
\newcommand{\actaa}{Acta Astron.} 
\newcommand{\araa}{ARA\&A} 
\newcommand{\aar}{A\&ARv} 
\newcommand{\aapr}{A\&ARv} 
\newcommand{\ab}{Astrobiol.} 
\newcommand{\aj}{AJ} 
\newcommand{\apj}{ApJ} 
\newcommand{\apjl}{ApJL} 
\newcommand{\apjs}{ApJSS} 
\newcommand{\ao}{Appl. Opt.} 
\newcommand{\apss}{Astro. \& Space Sci.} 
\newcommand{\aap}{A\&A} 
\newcommand{\aaps}{A\&AS.} 
\newcommand{\baas}{Bull. Am. Astron. Soc.} 
\newcommand{\caa}{Chinese A\&A} 
\newcommand{\cjaa}{Chinese J. A\&A} 
\newcommand{\cqg}{Class. Quantum Gravity} 
\newcommand{\gal}{Galaxies} 
\newcommand{\gca}{Geo. Cosmo. Acta} 
\newcommand{\icarus}{Icarus} 
\newcommand{\jcap}{JCAP} 
\newcommand{\jgr}{J. Geophys. Res.} 
\newcommand{\jgrp}{J. Geophys. Res. Planets} 
\newcommand{\jqsrt}{J. Quant. Spectrosc. Radiat. Transf.} 
\newcommand{\memsai}{Mem. SAIt} 
\newcommand{\mnras}{MNRAS} 
\newcommand{\nat}{Nature} 
\newcommand{\nastro}{Nat. Astron.} 
\newcommand{\ncomms}{Nat. Commun.} 
\newcommand{\nphys}{Nat. Phys.} 
\newcommand{\na}{New Astron.} 
\newcommand{\nar}{New Astron. Rev.} 
\newcommand{\physrep}{Phys. Rep.} 
\newcommand{\pra}{Phys. Rev. A} 
\newcommand{\prb}{Phys. Rev. B} 
\newcommand{\prc}{Phys. Rev. C} 
\newcommand{\prd}{Phys. Rev. D} 
\newcommand{\pre}{Phys. Rev. E} 
\newcommand{\prx}{Phys. Rev. X} 
\newcommand{\prl}{Phys. Rev. Let.} 
\newcommand{\psj}{Planet. Sci. J.} 
\newcommand{\planss}{Planet. Space Sci.} 
\newcommand{\pnas}{Proc. Natl Acad. Sci. USA} 
\newcommand{\procspie}{Proc. SPIE} 
\newcommand{\pasa}{PASA} 
\newcommand{\pasj}{PASJ} 
\newcommand{\pasp}{PASP} 
\newcommand{\rmxaa}{RMXAA} 
\newcommand{\sci}{Science} 
\newcommand{\sciadv}{Sci. Adv.} 
\newcommand{\solphys}{Sol. Phys.} 
\newcommand{\sovast}{Soviet Ast.} 
\newcommand{\ssr}{Space Sci. Rev.} 
\newcommand{\uni}{Universe} 

\setlength{\bibsep}{0.0pt}  
\maketitle

\section{Introduction}

The magnetic field is crucial to understanding a wide range of astrophysical processes such as star formation~\citep{Pattle+23}, galactic evolution~\citep{Marinacci+16}, and formation and evolution of cosmic large-scale structures~\citep{Subramanian+16}. How magnetic fields originate and evolve is an unsolved fundamental question~\citep{Brandenburg+23}. To achieve a breakthrough in answering the question, measurements of magnetic fields from the $\sim$pc scale in the interstellar medium (ISM) to the $\sim$Mpc scale in cosmic webs are required, to which SKA can make a unique major contribution.  

The prevailing theory for magnetic field evolution is the dynamo mechanism which converts kinetic energy to magnetic energy~\citep[see][for a review]{Brandenburg+23}. Dynamos have been proven to work over a tremendous range of scales from stars such as the Sun to clusters of galaxies. The mean-field dynamo produces the large-scale field, and the fluctuation dynamo produces the small-scale field. These dynamos co-exist in galaxies with the fluctuation dynamo thought to provide additional seed field for the mean-field dynamo~\citep[see][for details]{Shukurov+22}. However, there are still unresolved questions on how a dynamo works in a galaxy, and advancement of theory, simulations, and observations are needed to reach the complete picture.  

Radio observations provide two unique probes of magnetic fields: polarization imaging that involves deriving polarized intensity and rotation measure (RM) of diffuse emission and RM of extragalactic sources. At centimeter wavelengths, radio emission  originates mainly from synchrotron radiation by relativistic electrons spiraling in a magnetic field, which is linearly polarized. When a linearly polarized wave propagates in the magnetized ISM, its polarization angle $\chi$ rotates, which is called Faraday rotation. This can be represented as $\chi(\lambda)=\chi_0 + {\rm RM}\lambda^2$, where $\chi(\lambda)$ is the polarization angle at wavelength $\lambda$, $\chi_0$ is the intrinsic polarization angle, and ${\rm RM}=0.81\int_{\rm source}^{\rm observer}n_eB_\parallel{\rm d}l$. Here, the integral is along the line of sight from the source to the observer, $n_e$ is the thermal electron density in cm$^{-3}$, $B_\parallel$ is the line of sight component of the magnetic field in $\mu$G, ${\rm d}l$ is the increment in path length in pc and RM is in rad~m$^{-2}$.   

Polarized emission is modulated by Faraday rotation effects along the line of sight and across the beam~\citep[e.g.][]{Sokoloff+98}, which produces abundant structures in the polarization image, as can be seen from single-dish~\citep[e.g.][]{Sun+25,Ordog+26} and interferometer~\citep[e.g.][]{Wieringa+93,Haverkorn+00,Gaenlser+01,Gaensler+25} observations of Galactic diffuse emission. The decoding of the polarization images sheds light onto the properties of small-scale turbulent magnetic fields~\citep{Gaensler+11}. The polarization structures are wavelength dependent, and broadband polarization observations are needed to recover the emission and Faraday rotation structures. 

Each RM measurement traces the magnetic field along a single line of sight. An RM grid consisting of a large number of RMs of background extragalactic sources allows us to probe the magnetic field in a foreground object. An RM is composed of an intrinsic contribution from within the extragalactic source that has a scatter of $\sim$6~rad~m$^{-2}$~\citep[e.g.][]{Schnitzeler+10} and contributions from foreground objects. The denser an RM grid is, the more accurately the coherent structure caused by the foreground objects can be determined, and the more accurately the structure of the magnetic field of the foreground objects can be measured. Broadband polarization observations are also needed to accurately determine RMs.

With modern telescopes capable of multichannel broadband polarization observations, which deliver $Q(\lambda^2)$ and $U(\lambda^2)$, the RM synthesis technique~\citep{Burn+66, Brentjens+05} can be applied to obtain $Q(\phi)$ and $U(\phi)$. The polarized emission is thus decomposed into components of different Faraday depths $\phi$, and $\phi(\vec{r})=0.81\int_{\vec{r}}^{\rm observer}n_e B_\parallel{\rm d}l$ with $\vec{r}$ being the position inside the source and the rest being the same as in the definition of RM. The Faraday spectrum~\citep[e.g.][]{Sun+15b}, $F(\phi)\equiv Q(\phi)+iU(\phi)$, reflects the physical properties of a polarized source. For a source without internal RM, and assuming that the foreground medium is spatially uniform, $|F(\phi)|$ approaches a $\delta$ function with the peak position corresponding to the RM. Due to bandwidth, $Q(\lambda^2)$ and $U(\lambda^2 )$ are sampled in a limited number of $\lambda^2$, resulting in a spread function in the $\phi$ domain, called the RM spread function (RMSF). The full width at half magnitude (FWHM) of the RMSF determines the resolution in the $\phi$ domain and is proportional to $1/\Delta\lambda^2$, where $\Delta\lambda^2$ is the maximum separation of $\lambda^2$. The uncertainty of RM measurement can be estimated as $\sigma_{\rm RM}={\rm FWHM}/2{\rm SNR}_P$~\citep[e.g.][]{Vanderwoude+24}, where ${\rm SNR}_P$ is the signal-to-noise ratio in polarized intensity. This indicates the importance of a broad coverage in $\lambda^2$ space in the precise determination of RMs.

The Milky Way Galaxy is an ideal laboratory for understanding magnetism, for which we would be able to derive the geometry and strength of the large-scale magnetic field and statistical properties, such as correlation scales and power spectra, of random small-scale field. It is challenging for external galaxies~\citep{Mao01.2026.SKA} since each observation is an integral through the entire depth of the galaxy. In contrast, we are located inside the Galaxy and have the advantage of providing sight lines covering all directions, and the sight line integrates from the position of the observer to the edge of the Galaxy, thereby covering only part of the Galaxy. In addition, auxiliary information such as pulsar RMs, starlight polarization, or associations with objects with known distances can provide some 3D spatial information.

Obtaining an all-sky RM grid of the highest density and precision and all-sky polarization images of the Galactic diffuse emission covering scales larger than $\sim10\arcsec$ is what SKA will be able to achieve by conducting a deep all-sky polarization survey. The RM grid and polarization images will reveal the structure of the Galactic magnetic field, which will constrain theory and simulations. This will revolutionize the study of Galactic magnetism and also advance studies of the magnetic field beyond the Milky Way Galaxy. 

\section{Current status}
\subsection{Open questions on the Galactic magnetic field}
The Galactic magnetic field consists of a large-scale component (also called a regular field) with a coherent scale larger than $\sim$~kpc and a small-scale component (also called a random or turbulent field) on scales smaller than $\sim 100$~pc. The large-scale component is composed of a disk field and a halo field. 

The Galactic magnetic field has been extensively studied using RMs of pulsars and extragalactic sources~\citep[e.g.][]{Han+18, Xu+22, Curtin+24} and the total intensity and polarized synchrotron emission~\citep[e.g.][]{Adam+16}. Progress in these studies has been reviewed by \citet{Haverkorn+15b}, \citet{Han+17}, and \citet{Brandenburg+23}. Efforts have been made to derive the magnetic field model to reproduce all relevant observations including RMs, total, and polarized intensity~\citep[see][for a review]{Jaffe+19}, which started from simple qualitative modeling~\citep{Sun+08} to sophisticated quantitative modeling involving Markov Chain Monte Carlo~\citep{Boulanger+18} and $\chi^2$ minimization~\citep{Unger+24}.   

Despite extensive studies, there are still outstanding questions on the Galactic magnetic field that have yet to be answered. 

\textbf{What is the configuration of the halo field: toroidal, poloidal, X-shaped, dipole, quadrupole, \ldots?} The large-scale asymmetric pattern of RM signs below and above the Galactic plane, as can be seen from RMs of pulsars and extragalactic sources~\citep{Han+99, Taylor+09, Hutschenreuter+22}, can be naturally interpreted with a toroidal field with an odd symmetry with respect to the plane. The toroidal field has been incorporated in modeling to reproduce RMs and synchrotron emission. However, the exact form and extent of the toroidal field are very uncertain. The scale length of the toroidal field in Galactic radius ranges from $\sim4$~kpc~\citep{Sun+08} to $\sim10$~kpc~\citep{Unger+24} and $\sim15$~kpc~\citep{Xu+22}, and the functional form of the field also varies between models. In addition to the toroidal field, a poloidal field has also been proposed, which was driven by the RMs towards the Galactic poles. The average RM of extragalactic sources was measured to be $\sim6$~rad~m$^{-2}$ toward the southern Galactic pole~\citep{Taylor+09, Mao+10}. Toward the northern Galactic pole, it is uncertain with measurements of $\sim3$~rad~m$^{-2}$~\citep{Taylor+09, Sun+15} and $\sim0$~rad~m$^{-2}$~\citep{Mao+10}. This suggests that a simple dipole field~\citep{Han+17}, resulting in opposite RM signs toward the poles, is probably not sufficient to explain the observations. What is appealing is that the addition of a toroidal field and a poloidal field yields an X-shaped field that is often observed for extragalaxies~\citep{Krause+20}. The divergence-free poloidal field~\citep{Ferriere+14} and the X-shaped field have been used~\citep{Unger+24}. Recently, \cite{Dickey+22} found that RMs follow a $\sin(2l)$ pattern in the northern Galactic hemisphere and a $\sin(l+\pi)$ pattern in the southern hemisphere, which requires a more complex model of the whole field including halo and disk components, such as the combination of a dipole field and a quadrupole field. Here, $l$ is the Galactic longitude. Alternatively, the observed RM patterns could be dominated by more local structures~\citep[e.g.][]{West+21, Maconi+25}. Separating local magnetic feature contributions from the truly global large-scale halo field shall be carefully considered.

\textbf{What is the configuration of the disk field: grand design magnetic spiral arms, following matter spiral arms, axisymmetric, one or more field reversals, reversals not perpendicular to the disk, \ldots?} The magnetic field in the Galactic plane has reversals~\citep[see][for a review]{Han+17}, which is an unusual feature, one that is not observed in many other galaxies. It is challenging to determine the disk field with RMs because of the field reversal and with synchrotron emission because of the integral of long path lengths along line of sight and strong depolarization. The field pattern can be best studied with pulsar RMs and dispersion measures (DMs), and their ratios eliminate the dependence on the electron density. By segmenting the pulsars into distance bins, \cite{Han+18} found that the disk field follows the matter spiral arms with reversals alternating between arm and inter-arm regions~\citep{Han+17}. Note that the distances for most of the pulsars with known RMs are derived from DMs based on the Galactic thermal electron density model by \citet{Yao+17}. The median fractional uncertain of the distance is 15\% with an rms of 150\%. The most recent model by \citet{Ocker+26} expects to improve the distance estimate by achieving a fractional uncertainty of 0.9\% with an rms of 30\%. The distance uncertainty leads to the uncertainty of the derived structure of magnetic field. More direct distance measurements, such as from the very long baseline interferometer (VLBI) parallax, and more independent distance measurements, such as from the HI absorption, are required to measure the Galactic magnetic field more accurately with pulsar RMs and DMs. As an example, \citet{Curtin+24} used pulsars with the best distance estimates available to study the magnetic field along the spiral arms. To account for only the RMs of extragalactic sources, only one field reversal is required~\citep{Sun+08, VanEck+11}. Recent studies further suggest that reversal of the disk field can occur across the Galactic mid-plane \citep{Ma+20,Oswald+25}. Based on the RMs of pulsar and extragalactic sources, \citet{VanEck+11} found that the disk field follows the matter spiral arms only for the inner Galaxy and is purely azimuthal for the outer Galaxy. The recent models by \citet{Unger+24} show that the disk field is present between the matter spiral arms, forming magnetic spiral arms. Making the modeling of the disk field even more challenging, \citet{Ordog+17} found that the reversal is diagonal crossing the Galactic plane and a 3D modeling of the disk field is required.

\textbf{Where is the transition from the halo field to the disk field?} For most of the Galactic magnetic field models so far, the disk field has an even symmetry with respect to the Galactic plane; the halo field has a mixture of symmetry. The disk field decreases with increasing height, whereas the halo field varies in the opposite direction. The scale height of the disk field is about 1~kpc, as used by \citet{Sun+08} in an exponential function and derived by \citet{Unger+24} in a logistic sigmoid function. The toroidal field peaks at a height of 2-3~kpc~\citep{Sun+08, Xu+22, Unger+24}. A focused study towards the Perseus spiral arm suggests that the dominating component can transition from the disk to the halo at a Galactic height of $\sim 540\,{\rm pc}$ \citep{Mao+12}. With the toroidal field in the opposite direction of the disk field below the plane, a transition with sign change of the magnetic field is expected at the height of about 1~kpc. Above the plane, a transition of the total magnetic field strength from decreasing to increasing is expected. The transition becomes complex when the halo field includes more components. How the transition of magnetic field is reflected in RMs and synchrotron emission is yet to be explored.  

\textbf{What are the structure and energy spectrum of the small-scale field: purely random, anisotropic, ordered, striated, Kolmogorov-like, broken power law, \ldots?} Two of the characteristics of the small-scale magnetic field: structure and energy spectrum, are not yet clear. If the field is completely random, it will contribute to total synchrotron emission and cause depolarization, but the net RM is expected to be zero. An ordered~\citep{Jaffe+10} or striated~\citep{Jansson+12} small-scale field, which can be caused by compression or shearing of the random field, does not contribute to the net RM \citep[but can cause small-scale RM fluctuations;][]{Ma+25} but generates polarized emission. It is possible to differentiate the isotropic and ordered random fields with observables of total and polarized intensity and RM. There is a scarcity of measurement of the energy spectrum of the random field. \citet{Minter+96} showed that the spectrum can be derived from the structure function of RMs, and measured the spectrum on the scale of 0.01-100~pc for a small Galactic area with RMs of 38 extragalactic sources, implying a transition from 2D to 3D Kolmogorov turbulence. \citet{Han+17} showed that the energy spectrum depends on the scale and becomes flatter with increasing scales. Recently, \cite{Ma+25} showed that there can be an additional energy component at $< 0.1\,{\rm pc}$ scales on the Galactic mid-plane, which can be attributed to the anisotropic turbulent (i.e.\ striated) magnetic field, or to stellar feedback processes. The structure functions of the RMs also manifest a dependence on latitude~\citep{Sun+09, Stil+11} and arm-interarm environment~\citep{Haverkorn+08}, which complicates the measurement of the energy spectrum. 

\textbf{How are the large-scale and small-scale fields connected?} The small-scale field is generated by the small-scale fluctuation dynamo which quickly amplifies the weak seed field to strengths comparable to turbulent kinetic energy and then saturates. The large-scale mean field dynamo then acts on the small-scale field and produces the large-scale field~\citep{Shukurov+22, Brandenburg+23}. The large-scale field can also be tangled and converted back to the small-scale field \citep{Seta+20}. Therefore, a connection in the energy spectra of small-scale and large-scale fields is expected, but it has not yet been established firmly from observations~\citep{Han+17}. The spatial correlation of the regular and random fields is uncertain. In the models, a homogeneous and isotropic random field \citep{Sun+08} or an ordered random field correlated with the regular field~\citep{Jaffe+10, Jansson+12, Unger+24} was used. Interestingly, a correlation was found between the mean RM and the standard deviation of RMs, suggesting a correlation between regular and random fields~\citep{Brown+01}. From simulations, \citet{Wu+09} confirmed a tight correlation between the regular field strength and the width of the RM distribution caused mainly by a purely random field. The strength of the random field is generally greater than that of the regular field, with a ratio of 2-4 in the solar neighborhood~\citep[e.g.][for a review]{Haverkorn+15b}. With this large random field, the magnetic field estimated from RM might not be accurate due to the correlation between thermal electron density and random fields, particularly for small scales around $\sim$100~pc~\citep{Beck+03, Seta+21}.

\subsection{RM grid surveys}

\begin{figure}[!htbp]
    \centering
    \includegraphics[width=0.99\columnwidth]{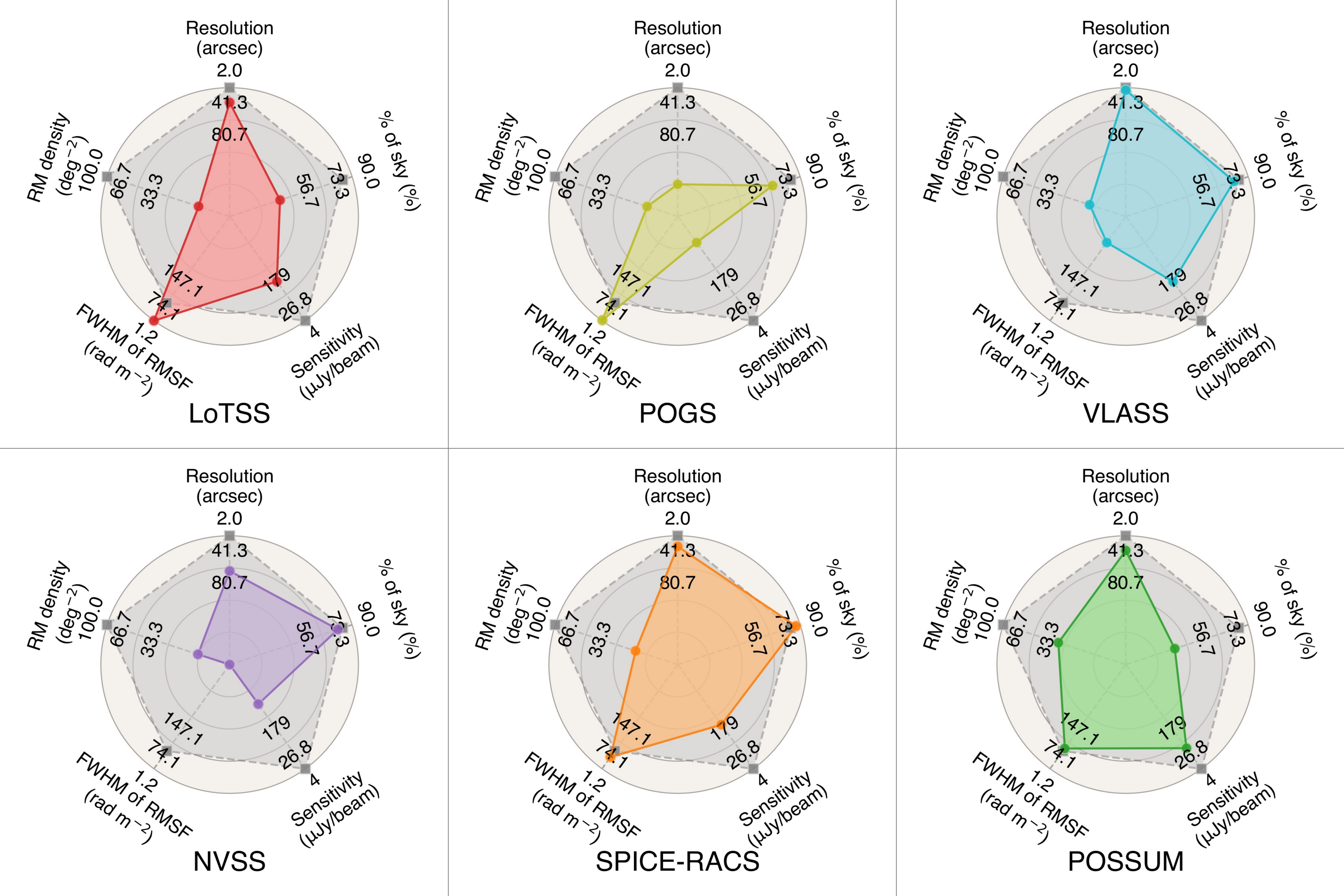}
    \caption{All-sky RM grid surveys: LoTSS, POGS, VLASS, NVSS, SPICE-RACS, and POSSUM, showing 5 key parameters: sky coverage, sensitivity, FWHM of the RMSF, RM density, and resolution. For comparison, an SKA-Mid survey discussed in Sect.~\ref{ska-mid} is shown in each panel. The sensitivity and FWHM of the RMSF are displayed on logarithmic scales. There is no RMSF for NVSS because RMs were derived from two frequencies not able to use RM synthesis.}
    \label{fig:rm_grid}
\end{figure}

The RM grid surveys target extragalactic sources and measure their RMs and polarized intensity. Interferometers are best suited to conduct these surveys as they can provide high angular resolution to allow for long integration time to overcome confusion and detect more sources. A summary of the completed and ongoing RM grid surveys was presented by \citet[][their Table 1]{Gaensler+25}, and most RM values available up to 2023 have been compiled by \citet{VanEck+23} into a consolidated catalog.

The key parameters characterizing an RM grid survey include sky coverage, frequency coverage, rms noise, resolution, RM density, and the FWHM of the RMSF. These parameters, except for frequency coverage that is encoded in the FWHM of the RMSF, are shown for several all-sky RM grid surveys in Fig.~\ref{fig:rm_grid}. To study the Galactic magnetic field, a dense RM grid with a large sky coverage is desired. 

The first RM catalog covering a large sky area with declination $\delta>-40\degr$ was produced by reprocessing the National Radio Astronomy Observatory (NRAO) Very Large Array (VLA) sky survey (NVSS) and has an RM density of about 1~deg$^{-2}$~\citep{Taylor+09}. Although these RMs were derived from the polarization angles at two very close frequency channels near 1.4~GHz and susceptible to $n\pi$ ambiguity \citep{Ma+19a}, they were proven to be statistically reliable compared to more precise measurements later~\citep[e.g.][]{Mao+12}. In the southern sky, the S-band Polarization All-Sky Survey~\citep[S-PASS,][]{Carretti+19}/Australia Telescope Compact Array (ATCA) survey \citep{Schnitzeler+19} is the first large-area RM survey, presenting RMs towards about 3,800 sightlines, though in some cases the low resolution caused additional Galactic components to be mixed in, mimicking complexity \citep[see e.g.][]{Ranchod+24}. These two RM grids were the primary data sources for constructing a Galactic RM sky by~\citet{Hutschenreuter+22} and modeling the large-scale magnetic field~\citep[e.g.][]{Jansson+12}. Recent low-frequency surveys, such as the POlarized GLEAM (GaLactic and Extragalactic All-sky MWA survey) Survey \citep[POGS,][]{Riseley+20} by the Murchison Widefield Array (MWA) and LOw Frequency ARray (LOFAR) Two-meter Sky Survey~\citep[LoTSS,][]{OSullivan+23}, also delivered all-sky RM grids with high RM accuracy, but the RM density is much lower due to strong depolarization at low frequencies.

The next big step towards a denser all-sky RM grid is the Spectra and Polarization In Cutouts of Extragalactic Sources - Rapid Australian SKA Pathfinder (ASKAP) Continuum Survey (SPICE-RACS) covering $\delta<+49\degr$. The first data release for a sky area of about 1300~deg$^2$ shows that the RM density is about 4~deg$^{-2}$~\citep{Thomson+23}, and the second data release for all-sky RMs shows that the RM density is about 7~deg$^{-2}$~\citep{Thomson+26}. The eventual plan for SPICE-RACS to combine the three ASKAP frequency bands in the range of 800-1800~MHz will deliver an expected RM density of about 10~deg$^{-2}$~\citep[][their Table~1]{Gaensler+25}, reaching an order of magnitude improvement compared to the NVSS RM grid. 

The ongoing VLA Sky Survey~\citep[VLASS,][]{Lacy+20} at 2-4~GHz (S band) will also deliver an all-sky RM grid with a density of about 6~deg$^{-2}$. The FWHM of the RMSF is approximately 220~rad~m$^{-2}$, which means a large uncertainty of RM measurement. This grid is ideal for studying the intrinsic properties of extragalactic sources, but will also provide unique sight lines through the Galaxy since there would be a population of sources that are only polarized at S band. 

The Polarization Sky Survey of the Universe’s Magnetism (POSSUM), which is being conducted by ASKAP, will make a big leap forward by providing an RM density of 30-50~deg$^{-2}$ covering approximately the southern 50\% of the sky~\citep{Gaensler+25}. The primary frequency range is 800-1088~MHz, with supplementary data at 1296-1440~MHz, and RMs are measured with RM synthesis, resulting in a median uncertainty of about 1~rad~m$^{-2}$. As demonstrated by \citet{Gaensler+25}, details of the RM structures can be revealed with the POSSUM RM grid, which were not recognized from the NVSS RM grid. The POSSUM survey is expected to be completed by the Middle of 2028.

\subsection{Diffuse polarized emission surveys}
Surveys targeting diffuse polarized emission from the Galaxy are usually conducted with single-dish telescopes to preserve the large-scale emission which is missed in observations with interferometers. Polarization surveys nowadays are usually observed with broadband multichannel receivers, which enables RM synthesis to derive Faraday depth and polarized intensity simultaneously.  

The Global Magneto-Ionic Medium Survey (GMIMS) is a major effort to study the diffuse polarized emission with RM synthesis, aiming to observe the full sky with telescopes in both hemispheres covering the frequency range of 300-1800~MHz. Four GMIMS component surveys have been completed with data released: the low-band north (350-1030~MHz) survey conducted with the Dominion Radio Astrophysical Observatory (DRAO) 15-m telescope \citep{Ordog+26},  the high-band (1280-1750~MHz) north ($-30\degr<\delta<+87\degr$) survey conducted by the  DRAO 26-m telescope~\citep{Wolleben+21}, the low-band (300-480~MHz) south ($-90\degr<\delta<+20\degr$) survey conducted by Murriyang, the Parkes 64-m telescope~\citep{Wolleben+19}, and the Southern Twenty-centimeter All-sky Polarization Survey \citep[STAPS,][]{Sun+25} conducted by Murriyang as the high-band (1.3-1.8~GHz) south ($-89\degr<\delta<0\degr$) component. Among these surveys, STAPS has the highest resolution of about $20\arcmin$. These surveys delivered frequency cubes: $I(\lambda^2)$, $Q(\lambda^2)$, and $U(\lambda^2)$ and Faraday depth cubes: $Q(\phi)$, $U(\phi)$, and $F(\phi)$. 

The Five-hundred-meter Aperture Spherical radio Telescope (FAST) is conducting the Commensal Radio Astronomy FAST Survey \citep[CRAFTS,][]{Li+18}, which observes pulsar, Galactic and extragalactic HI, and continuum simultaneously. The frequency coverage is 1.0-1.5~GHz, the planned sky area is $-14\degr<\delta<+66\degr$, and the resolution is $3\arcmin$-$4\arcmin$. \citet{Sun+22} demonstrated the great polarization capability of FAST by showing the RM synthesis results of the Cygnus Loop supernova remnant. The polarization processing of CRAFTS is currently underway~(Sun et al. in prep.). 

POSSUM and the total intensity survey of ASKAP, the Evolutionary Map of the Universe~\citep[EMU,][]{Hopkins+25}, can preserve the emission up to an angular scale of about $30\arcmin$ at 944~MHz. In order to provide the larger scale emission for POSSUM and EMU and fill the frequency gap of the GMIMS survey to better determine RMs of diffuse emission, the POSSUM-EMU-GMIMS All Stokes UWL Survey (PEGASUS) has been proposed (Carretti et al., Murriyang project ID P1123) and is currently being observed with Murriyang using the Ultra-Wide-bandwidth Low-frequency receiver (UWL). PEGASUS covers the frequency range of 700-1440~MHz overlapped with STAPS and the sky area of $-90\degr<\delta<+20\degr$. The observations are expected to be completed by July 2026, and the data processing is expected to be finished by the middle of 2028, roughly the same as POSSUM.

\section{An SKA-Mid all-sky polarization survey}
\label{ska-mid}

SKA will be able to advance the understanding of the Galactic magnetic field through an all-sky polarization survey, which delivers an RM grid that contains millions of RMs and polarized intensity and Faraday depth spectra of the Galactic diffuse emission that cover angular scales from arcseconds to degrees after combining with single-dish observations.   

The RM grid has been and will continue to be the highest priority of SKA magnetism over the years of developing the science case~\citep{Beck+04, Haverkorn+15, Johnston+15, Heald+20}. An all-sky polarization survey with SKA-Mid Band 2 to deliver this RM grid has been elaborated by \citet{Heald+20}: sky coverage of 30,\,000 deg$^2$, frequency range of 0.95-1.76~GHz, resolution of $2\arcsec$, and rms sensitivity of 4~$\mu$Jy~beam$^{-1}$. The nominal integration time for each pointing is 15~min, and it will take about 2.5~yr total time, assuming observations are conducted at night and including overheads, to complete the survey that contains about 30,\,000 pointings~\citep{Heald+20}. This SKA-Mid survey compared to other surveys is shown in Fig.~\ref{fig:rm_grid}. The survey improves significantly in RM density, resolution, and sensitivity.

\subsection{Perspective of AA4 and AA*}
The SKA will be delivered in stages (SKAO-TEL-0002299). There will be a planned pause after AA* and the delivery of AA4 will depend on the funding. AA* contains 144 dishes, including 64 MeerKAT 13.5-m dishes and 80 SKA 15-m dishes. AA4 will add another 53 15-m dishes, reaching a total number of 197. 

According to the most recent SKA Observatory (SKAO) sensitivity calculator\footnote{https://sensitivity-calculator.skao.int/}, the rms sensitivity with AA* reaches about 3.4~$\mu$Jy~beam$^{-1}$ with Briggs weighting and robust of 0 for an integration time of 15~min. For AA4, it takes about 7~min of integration time per pointing to reach a similar sensitivity, which is about half the integration time with AA*. To complete the all-sky RM grid survey, it will take about 1.2~yr total time. If we keep the same integration time, the sensitivity will reach 2.35~$\mu$Jy~beam$^{-1}$, and the density of the RM grid increases by a factor of approximately 1.2, as discussed below.

The resolution with AA4 is about $0\farcs8$. At this resolution, most of the extragalactic sources will probably be partially or totally resolved. There may be potential risks here. For a source that is compact at $2\arcsec$ resolution, the specific intensity in polarization becomes lower if resolved at $0\farcs8$ resolution, and it might not be detected with the same sensitivity. This will cause a decrease of the source count. When most of the sources are resolved, the intrinsic RM structure of the sources becomes important, which poses a challenge when using the RM grid to study the foreground magnetic field.    

The maximum baseline is about 23~km for AA* after excluding the remote dish SKA-008, which leads to a resolution of $2\arcsec$ specified for the RM grid survey. For AA4, there will be 43 more dishes located within the maximum baseline of AA*. If the resolution of $2\arcsec$ is kept by excluding long baselines, the collecting area for AA4 is about a factor 1.3 larger than AA* and the total time it takes for the survey is about 1.9~yr. 

We strongly advocate starting the RM grid survey with AA*. The SKA-Mid RM grid survey is ideally conducted with AA4 as it takes less time or delivers more RMs provided that the high resolution resulting in resolving sources is not concerning. However, since AA* has already delayed, AA4 is likely to be further delayed given the funding uncertainty.  When AA4 is ready, very deep observations of small sky areas can be conducted because the higher resolution, meaning a lower confusion limit, allows for a longer integration time.  

\subsection{Survey specifications}

\textbf{RM grid density.} Polarization source counts at $\sim10$~$\mu$Jy level has been studied only recently with some deep fields. The total number of polarized sources per deg$^2$ with polarized flux density larger than $P$ can be represented as $N(>P)=N_0(P/P_0)^\alpha$, where $P_0=30$~$\mu$Jy. For $P<1$~mJy, \citet{Rudnick+14} found $\alpha=-0.6$ and $N_0=45$~(68) with a resolution of $1\farcs6$ ($10\arcsec$) based on Jansky VLA observations. Similarly, \citet{Berger+25} obtained $\alpha=-0.54$ and $N_0=66.5$ using deep field observations with the Westerbork Synthesis Radio Telescope (WSRT) at a resolution of about $15\arcsec$. The expected number of sources per deg$^2$ would be about 127 and 92 for a $P$ threshold of 9~$\mu$Jy and 16.4~$\mu$Jy, respectively. However, the source number was found to be about 160~deg$^{-2}$ with a resolution of $13\arcsec$~\citep{Loi+25} and about 138~deg$^{-2}$ with a resolution of $18\arcsec$~\citep{Taylor+24} corresponding to these two thresholds, higher than expected by a factor 1.3-1.5. Assuming a $5\sigma$ threshold of peak polarized intensity to measure RMs, the density of the SKA-Mid RM grid is expected to be 108-124 (117-136)~deg$^{-2}$ for a $5\sigma$ threshold of 20 (17)~$\mu$Jy, slightly higher than the estimate by \citet{Heald+20}. The density of the RM grid would be a factor of 3 higher than POSSUM~\citep{Gaensler+25}. In the following, we quote a fiducial number of 100~deg$^{-2}$ when discussing science with the RM grid. 

\textbf{Sky coverage.} As emphasized by \citet{Heald+20}, a large sky coverage, covering the observable all-sky area of about 30,\,000 deg$^2$, is key to the success of the survey. We advocate for the survey to be wide instead of deep for the following reasons. (1) The Galactic magnetic field has a complicated structure that produces longitude- and latitude-dependent RM variations. A dense all-sky RM grid is essential to reveal these variations and differentiate the models of magnetic field. The discovery of sinusoidal RM patterns toward the Galactic northern hemisphere~\citep{Dickey+22} and the RM gradient across a diagonal line instead of longitude or latitude~\citep{Ordog+17} with more RM data are good examples. (2) Comparison of RMs of extragalactic sources with Faraday depths of diffuse emission is essential to disentangle emission structures along the line of sight, as the ratio reveals how thermal gas and nonthermal gas are mixed~\citep{Sokoloff+98,Erceg+22}. This will identify whether the RM features are caused by discrete objects or by the Galactic magnetic field. Therefore, an all-sky RM grid is required to match all-sky images of polarized emission. (3) A dense all-sky RM grid is needed to build an accurate RM foreground model of the Milky Way Galaxy, which is crucial to determine RMs of extragalactic objects such as fast radio bursts (FRBs), nearby galaxies, and clusters of galaxies. The current model by \citet{Hutschenreuter+22} is based on RM measurements from various observations containing different systematics. With the dense all-sky RM grid, the intrinsic contributions of the sources can be better removed and the coherent RM structures can be better recognized, so that an accurate RM model can be achieved. (4) Besides the magnetic field in the Milky Way Galaxy, magnetic fields in the sources themselves, in nearby galaxies, and in other large-scale objects will also be studied. It is important to have all-sky polarization observations to increase the samples to overcome statistical errors due to the small sample size.         

\textbf{Angular resolution.} The resolution of $2\arcsec$ is optimized for source cross-identification with optical observations~\citep{Prandoni+15, Heald+20}, and is about 10 times better than POSSUM. At this resolution, a large number of extragalactic sources will be resolved into multiple components. These components provide close sight lines, and RM variations across these sight lines will allow us to probe the turbulent magnetic field to the scale of tens of arcseconds~\citep[e.g.][]{Minter+96,Ma01.2026.SKA}. This will also help eliminate the Galactic contribution when studying the magnetic field in the intergalactic medium with adjacent pairs of RMs~\citep{Vernstrom+19}.

\textbf{Frequency coverage.} The SKA-Mid Band 2 has a frequency range of 950-1760~MHz, which is ideal for RM measurements using RM synthesis. According to the SKAO sensitivity calculator, the central frequency is 1.31~GHz with a bandwidth of 720~MHz when including the 64 MeerKAT dishes, and the corresponding FWHM of the RMSF is 51.43~rad~m$^{-2}$. For a signal-to-noise ratio of 5 as a threshold for RM determination, the RM uncertainty is better than about 5~rad~m$^{-2}$. At higher frequency, the source count will decrease as a result of the steep spectra of extragalactic sources, and the FWHM of the RMSF will increase, resulting in a larger RM uncertainty. At lower frequency, such as the frequency range of 350-1050~MHz for SKA-Mid Band 1, the FWHM of the RMSF is about 6~rad~m$^{-2}$, much narrower than that for SKA-Mid Band 2. However, strong depolarization will occur at the low frequency end, which largely reduces the number of detectable polarized sources. The broad band will also allow for a depolarization analysis of the sources to study their environment. 

\begin{figure}[!htbp]
    \centering
    \includegraphics[width=0.98\columnwidth]{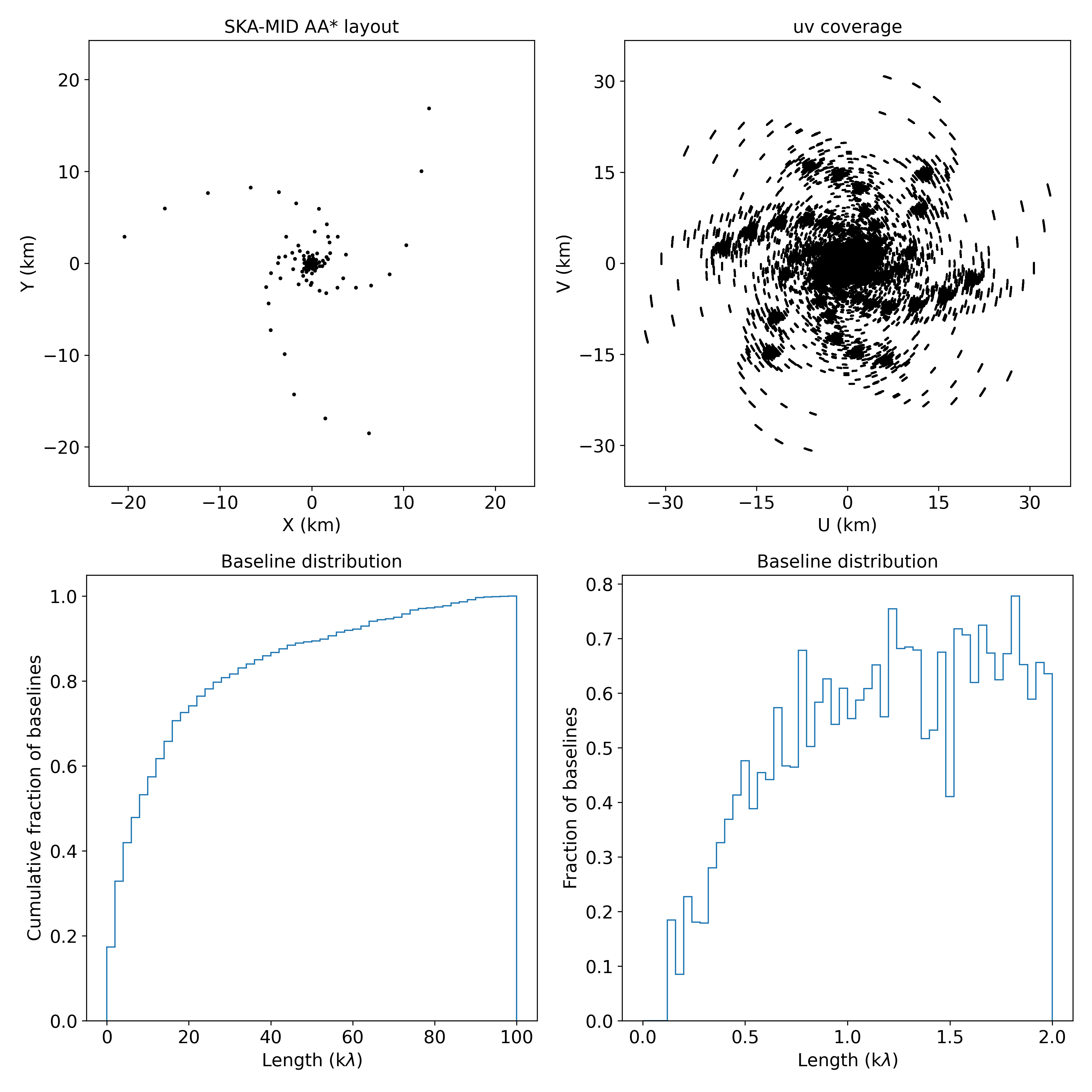}
    \caption{Antenna layout for AA* and $uv$ coverage and distributions for a 15~min observation of a source at $\delta=-60\degr$ at the central frequency of 1.31~GHz of SKA-Mid Band 2. The remote dish SKA-008 was excluded.}
    \label{fig:simulation}
\end{figure}

\textbf{\bm{$uv$} coverage.} We simulated a 15~min observation of a source at $\delta=-60\degr$ and plotted the antenna layout for AA* and the $uv$ distributions using the \verb|ska_ost_array_config|\footnote{https://gitlab.com/ska-telescope/ost/ska-ost-array-config} package. Note that the simulations results are similar for AA4 if the $2\arcsec$ resolution is used. The frequency is 1.31~GHz which is the central frequency of SKA-Mid Band 2 if including MeerKAT dishes. The results are shown in Fig.~\ref{fig:simulation}. About 20\% of the baselines is less than about 2~k$\lambda$ and the minimum baseline is about 0.15~k$\lambda$, corresponding to a maximum angular scale of about $22\arcmin$ for large-scale emission. This overlaps well with single-dish polarization surveys by Murriyang and FAST, and these data can be combined to obtain emission that covers all scales above the resolution.  

\textbf{Readiness.} All-sky RM grid surveys such as POSSUM with ASKAP~\citep{Gaensler+25} and the deep field RM grid surveys with MeerKAT~\citep{Taylor+24, Loi+25}, have developed polarization calibration and imaging techniques that can be used for the SKA-Mid polarization survey. The strong sources from POSSUM can be used as a sky model to facilitate the calibration of the SKA survey. The RM-Tools package~\citep{Purcell+20, vaneck+26} is versatile for RM determination and complexity diagnosis. 

\section{Galactic magnetic field with an SKA-Mid polarization survey}

An SKA-Mid polarization survey will deliver an all-sky dense RM grid and total intensity and polarization images of the Galactic diffuse emission. The density of the RM grid is about two orders of magnitude higher than NVSS, about three times that of POSSUM, and roughly the same as that of the MeerKAT deep field.  

\subsection{RM grid}

\begin{figure}[!htbp]
    \centering
    \includegraphics[width=\columnwidth]{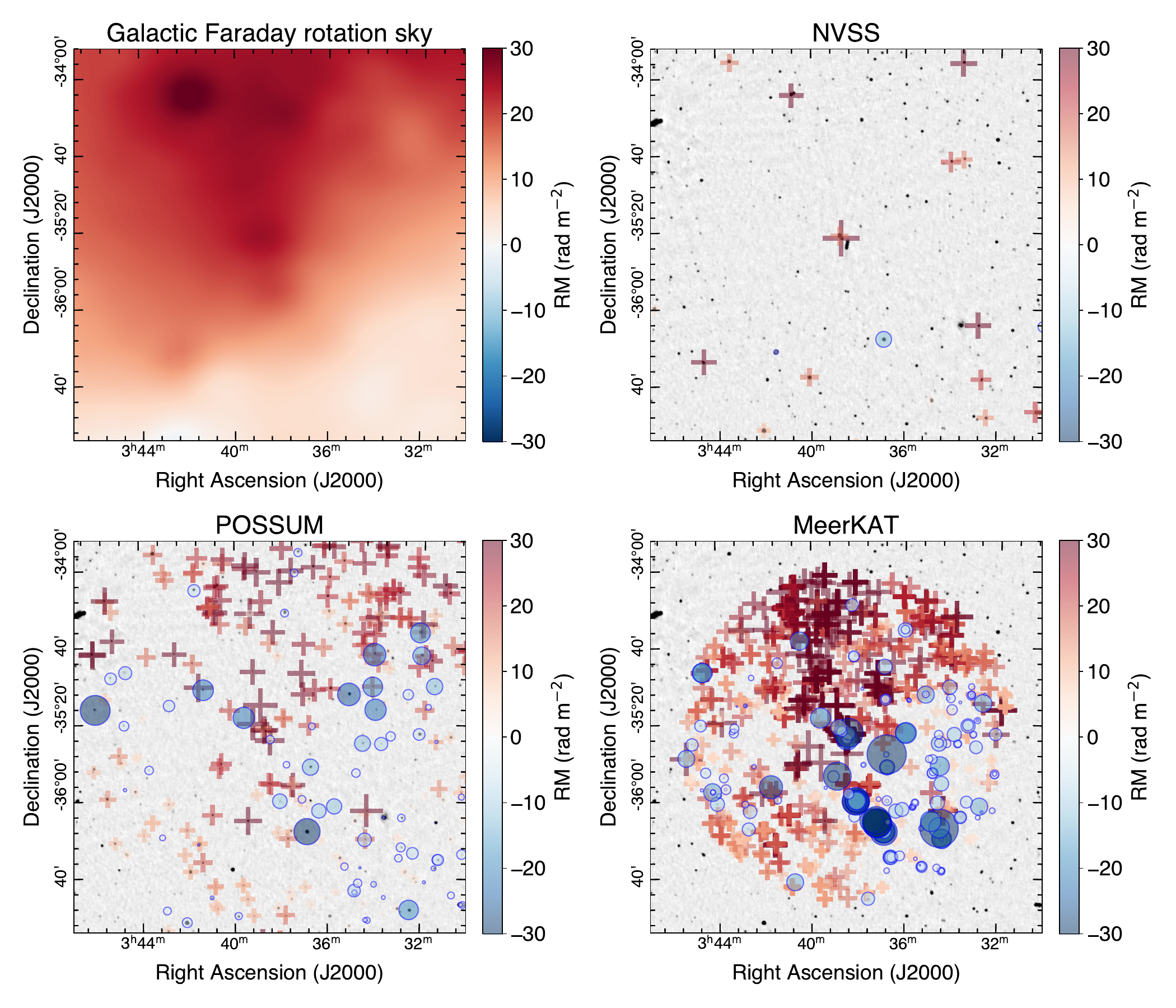}
    \caption{Image of the Galactic RM sky~\citep{Hutschenreuter+22} and RMs of extragalactic sources from NVSS~\citep{Taylor+09}, POSSUM~\citep{Anderson+21}, and MeerKAT~\citep{Loi+25} overlaid on the NVSS total intensity image at 1.4 GHz.}
    \label{fig:rm-comp}
\end{figure}

We show the Galactic RM sky~\citep{Hutschenreuter+22} constructed mainly from the NVSS and S-PASS/ATCA surrveys, and RMs of extragalactic sources from NVSS~\citep{Taylor+09}, POSSUM~\citep{Anderson+21}, and MeerKAT~\citep{Loi+25} in Fig.~\ref{fig:rm-comp}. The $3\degr\times3\degr$ area is centered on the Fornax cluster of galaxies. 

It can be clearly seen that the RM sky and NVSS RMs cannot represent the true RM distribution of the Galaxy. The conspicuous pattern of a negative RM strip from the northeast to the southwest through the field, as seen with POSSUM, is missed in NVSS. It appears that the negative RM pattern extends beyond the field. With the densest RM grid of MeerKAT, strips of enhanced positive and negative RMs can be identified toward the north and southwest. These strips also appear to extend beyond the field, probably representing a foreground feature. 

The all-sky RM grid from the SKA-Mid polarization survey will reveal the accurate Galactic foreground RM. With a density of about 100~deg$^{-2}$, we can reach the scale of $\sim10\arcmin$. RMs of extragalactic sources consist of an intrinsic component from magneto-ionic gas inside the source and an extrinsic component from gas between source and observer. For the latter, the contribution from the Galactic medium is dominant. The intrinsic RM components are not correlated. In contrast, the Galactic RM components are correlated over a scale depending on the Galactic magnetic field. To estimate the Galactic RM component, the widely used method is to do a running average~\citep[e.g.][]{Unger+24}. For a low density RM grid, the method will be biased toward intrinsic RM component or RM from Galactic objects. The problem will be alleviated with a denser RM grid. The running average is certainly not sufficient to capture the spatial RM variation. More advanced method by \citet{Hutschenreuter+22} has been used to construct a Faraday depth sky of the Milky Way. There is work to be done to develop robust method to properly extract the Galactic RM component from a dense RM grid. \citet{Khadir+24} tested interpolation algorithms with an RM grid of $\sim40$~deg$^{-2}$ to recover patchy or filamentary structures, and found that the Bayesian RM sky method worked the best. Future studies with different underlying foreground RM structures and a denser RM grid are needed to develop interpolation techniques. 

\subsection{Diffuse emission}
\begin{figure}[!htbp]
    \centering
    \includegraphics[width=\columnwidth]{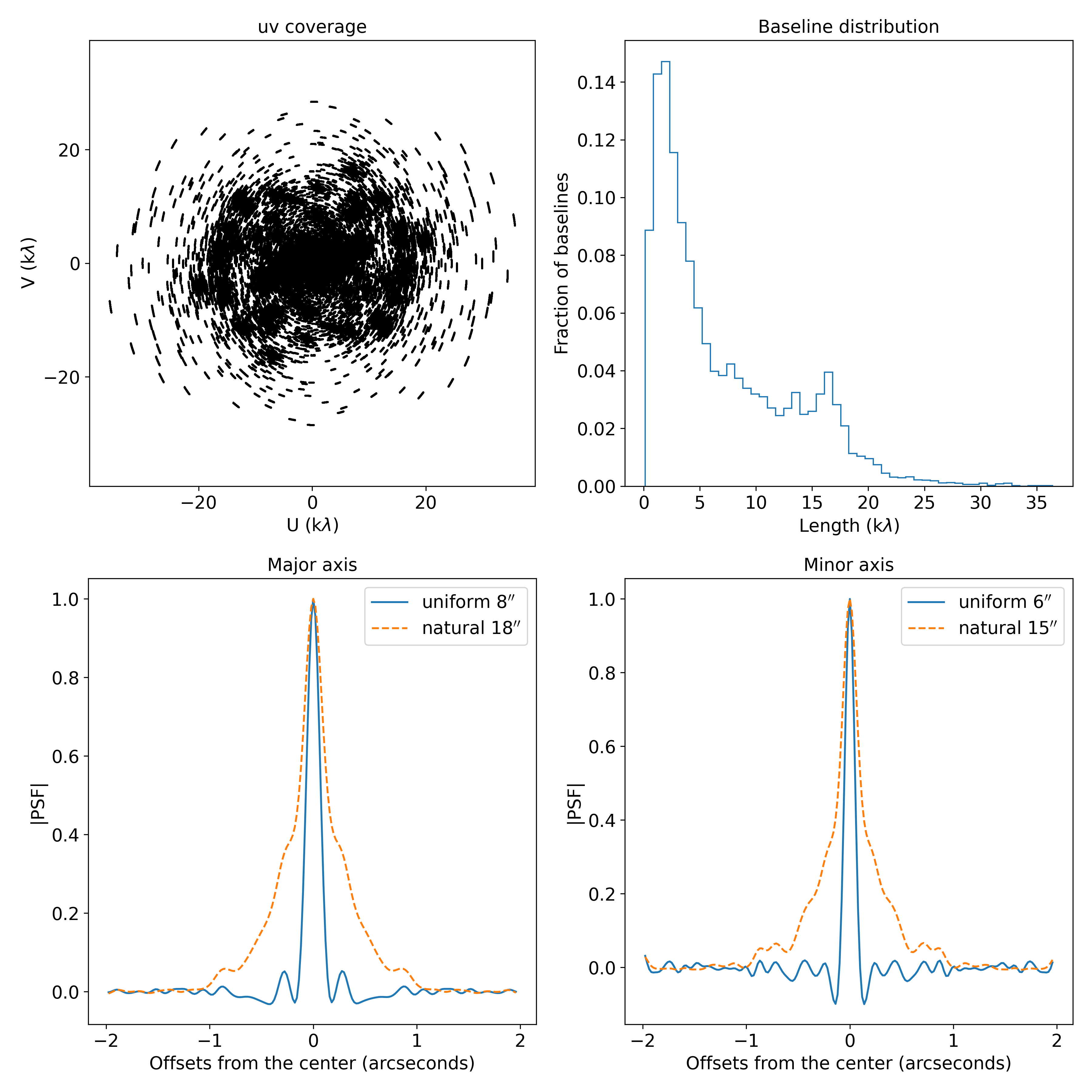}
    \caption{$uv$ coverage, baseline distribution, and PSFs along the major and minor axes for uniform and natural weightings for a simulated 15~min observation with AA*, only using dishes within 5~km of baselines. The PSF sizes are displayed in the two bottom panels.}
    \label{fig:beam}
\end{figure}

Most of the dishes, 123 out of 144, are located within a maximum baseline of 5~km for AA* and are well suited to observe diffuse emission. We simulated an observation of a source at $\delta=-60\degr$ for 15~min, only using dishes within 5~km. The $uv$ coverage, the baseline distribution, and the point spread functions (PSFs) along the major and minor axes for uniform and natural weightings are shown in Fig.~\ref{fig:beam}. The PSF sizes from the natural weighting are only about a factor 2 larger than those from the uniform weighting, indicating a great surface brightness sensitivity that is needed for imaging the diffuse emission. If we taper baselines in $uv$ beyond 20~k$\lambda$, a well-shaped synthesized beam is expected and the resolution is about $10\arcsec$ with a Briggs weighting of robust 0.  

\begin{figure}[!htbp]
    \centering
    \includegraphics[width=\columnwidth]{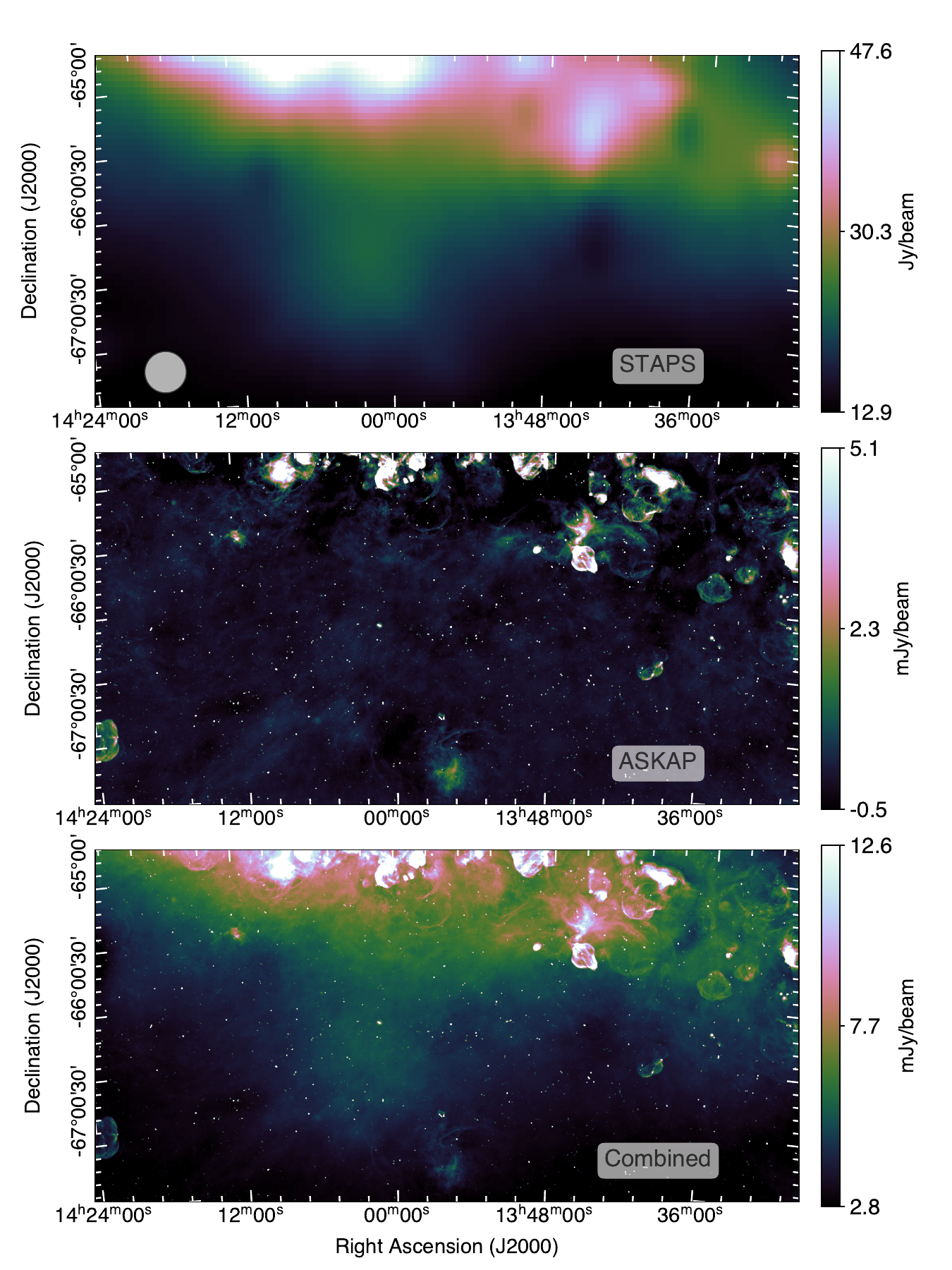}
    \caption{Total intensity images from STAPS~\citep{Sun+25}, ASKAP EMU~\citep{Hopkins+25}, and the two combined.}
    \label{fig:combine}
\end{figure}

The combination of interferometer and single-dish data is essential to retain the large-scale emission. There are methods working on visibility data~\citep{Rau+19,Ordog+25} and images~\citep{Jiao+22}, and a review was presented by \citet{Plunkett+23}. The methods are still being developed. We combined the EMU and STAPS data with \verb|CASA feather| and the results are shown in Fig.~\ref{fig:combine}. Note that the frequencies of the two surveys do not match, but are very close, and we extrapolated STAPS data assuming a spectral index for the combination. As can be seen, the combined image shows structures of scales from the EMU resolution of $18\arcsec$ to larger. The combination in polarization is more complex, and $Q$ and $U$ shall be combined separately for each individual channel before RM synthesis. For both total intensity and polarized intensity, joint deconvolution of visibilities, such as \verb|TP2VIS|~\citep{Koda+19} developed for ALMA, likely works better, but further tests are needed.  

For the frequency range of 950-1760~MHz for SKA-Mid Band 2, the maximum detectable Faraday depth scale is about 108~rad~m$^{-2}$~\citep{Brentjens+05}. Ideally, the FWHM of the RMSF is required to be much smaller than the maximum scale to properly recover the structures of the Galactic diffuse polarized emission that is distributed over extended Faraday depth. The SKA-Mid Band 1 is thus favored since it yields a much narrower FWHM of the RMSF. However, the maximum Faraday depth scale from SKA-Mid Band 1 is also much smaller. This means that polarized emission from different regions are detected for different frequency ranges. Observations with SKA-Mid Band 1 will help understand polarized structures that are closer to the observer. A combination of observations spanning a wide frequency range will deliver a 3D structure of the interstellar medium. 

The total intensity and polarization images from the SKA-Mid polarization survey in combination with single-dish observations will reveal the emission structures of the Galactic interstellar medium down to a scale of about $10\arcsec$.  

\subsection{Galactic magnetic field}

Breakthroughs in studying the Galactic magnetic field are expected with the SKA-Mid RM grid and images of diffuse emission. 

\textbf{Large-scale field.} With the dense RM grid, the RM caused by the Galactic magnetic field can be accurately extracted. An RM grid of 100~deg$^{-2}$ will be able to identify RM caused by Galactic objects such as bubbles and supernova remnants with angular size larger than $\sim10\arcmin$, combining multiwavelength observations probing these objects. The accurate RM pattern of the Galaxy can thus be revealed, such as the RM variation versus longitude and latitude. The RM sign changes in the halo, indicating field reversals that are caused by the transition of disk field to halo field, can also be identified. Compared to the RM grid used so far for magnetic field modeling, the RM grid density improves by about two orders of magnitude. This will allow us to precisely determine the parameters that describe the Galactic magnetic field, using optimization techniques such as MCMC. The ratio of RM of extragalactic sources to that of diffuse emission is an indicator of how the synchrotron-emitting gas and the Faraday-rotating gas are mixed, which enables modeling of RMs and emission together.  

\textbf{Small-scale field.} The structure function of the RMs of unresolved extragalactic sources can be derived down to a scale of $\sim10\arcmin$, or down to an even smaller scale of $\sim1\arcmin$ since the sources can be closer to each other. If a source is resolved into multiple components, RM of each component will provide a probe of magnetic field along line of sight. This will be equivalent to a denser RM grid. The structure function of the RMs of the components of resolved sources is down to a scale of approximately $2\arcsec$, namely the resolution of the observations. However, in this case, the intrinsic contributions of RM are also correlated, and how to decompose RM into intrinsic and extrinsic components need to be further explored. Based on the Faraday complexity, RM structures of scales less than the beam size can also be inferred~\citep{Ma01.2026.SKA}. The power spectrum of intensity and RM of the diffuse emission can be evaluated on scales of about $10\arcsec$ upward after combination with single-dish observations. The power spectrum of random fields can then be inferred from these structure functions and power spectra on scales from $\sim10\arcsec$ to tens of degrees of the size of observed sky area, spanning three orders of magnitude. Power spectrum analysis can also be applied to different sky areas and different Faraday depths to obtain 3D properties of random fields.

\textbf{Connecting large-scale and small-scale fields.} Applying techniques such as gradient to total intensity and polarized intensity images will be able to derive the Alfv\'enic Mach number which is an indicator of the ratio of random to regular field~\citep{Lazarian+24}. With the dense RM grid, it is feasible to compare the standard deviation and average of RMs, which reflects the ratio of random to regular field and the scale of the random field. With total intensity and polarization images that both retain the scales from $10\arcsec$ upward, an accurate fractional polarization can be derived, from which the ratio of random to regular field can be extracted. 

\textbf{Magnetic field in discrete objects.} With the dense RM grid, magnetic fields in the foreground objects such as HII regions, HI cloud and supernova remnants (SNRs), whose angular scales larger than $\sim10\arcmin$ can be measured. Further exploration of the three-dimensional magnetic structure of discrete Galactic objects is discussed by~\citet{Tahani01.2026.SKA}. With the high-resolution and high-sensitivity images, it is expected that more SNRs of smaller angular sizes will be discovered, and it will be possible to measure their magnetic fields. This will also help solve the missing SNR problem~\citep{Ball+23, Ball+25}. By cross matching the RM grid with other surveys such as HI, H$\alpha$, and X-ray, it is anticipated that anomalous RM structures, such as the Faraday screen with very weak electron density but large regular magnetic field~\citep[e.g.][]{Sun+07}, will be discovered.

\textbf{Origin and evolution of the Galactic magnetic field.} One consequence of dynamo theory is the presence of helical field~\citep[e.g.][]{Brandenburg+23}. Magnetic helicity can be measured using the relationship between fractional polarization and RM~\citep[e.g.][]{West+20}. The dense RM grid and high resolution polarization image provide us the best opportunity to measure helicity, which will confirm dynamo as the origin of the Galactic magnetic field. With the complete picture of the Galactic magnetic field including the structure and the energy spectrum to be unveiled by the SKA-Mid polarization survey, how the dynamo works in the Galaxy will be understood. 

\subsection{Synergy with dust polarization and impact on other researches}

\textbf{Dust Polarization, dust physics, and dust properties.} Dust polarization provides an independent probe of the 3D Galactic magnetic field through the combination of the polarization angle and the polarization degree \citep{HoangTruong2024,TruongHoang2025}. It is possible to constrain the magnetic field structure combining synchrotron emission and dust polarization~\citep{Pelgrims+21}. However, the fundamental physics of grain alignment and how grain alignment efficiency varies with local conditions are still not fully understood~\citep[see][for a review]{Tram+22}. The dust polarization degree provides crucial constraints on the grain alignment physics and dust properties (e.g., size, shape, composition); yet, there is a degeneracy between grain alignment, dust properties, and the magnetic field in the dust polarization. This degeneracy can be disentangled only when 3D magnetic fields are constrained. The combination of plane-of-sky component of the magnetic field traced by dust polarization angle with line-of-sight component of the magnetic field traced by Faraday rotation from the SKA-Mid polarization survey will allow us to infer 3D magnetic fields of discrete objects~\citep{Tahani01.2026.SKA}. The inferred 3D magnetic fields will be used to model dust polarization based on the modern Radiative Torque (RAT) alignment theory \citep{HoangLaz2016}, which will be compared with the observed starlight and thermal dust polarization to constrain grain alignment mechanisms and dust properties in the different environments, from large to small scales.

\textbf{Ultra-high-energy Cosmic rays (UHECRs).} The origin of UHECRs (energy$\gtrsim10^{17}$~eV) is still a mystery. One of the keys to understanding the origin is to backtrack in which direction they enter the Galaxy based on the ground-based observations. The UHECRs are deflected by the Galactic magnetic field. As shown by \citet{Unger+24}, the backtrack direction differs significantly from the current models of the Galactic magnetic field. With the Galactic magnetic field model derived from the SKA-Mid polarization survey, the direction will be determined, which will advance the studies of UHECRs.   

\textbf{Low energy cosmic-ray flux and spectrum.} Cosmic-ray (CR) electrons with energies of 0.1-10~GeV produce the observed radio emission with frequencies up to GHz through synchrotron radiation. However, because of solar modulation, the flux and spectrum of these CR electrons are difficult to measure. The spectral index of the total intensity provides a way to derive these quantities~\citep{Bracco+24}, which also probes the plane-of-sky component of the magnetic field. With the broadband all-sky total intensity image, we will derive spectral index and thus measure the energy flux and spectrum of these low energy CR electrons, which can help us understand how they propagate in the Galaxy.   

\textbf{Accurate foreground RM model.} With the dense RM grid, an accurate foreground RM model of the Galaxy will be established. The dense RM grid will also be used to study the magnetic field in extragalactic objects, such as nearby galaxies, clusters of galaxies, and large-scale structures in cosmology. For these studies, it is essential to remove the dominant contribution from the Galaxy, which will benefit from the accurate foreground RM model. The foreground model is also required to obtain intrinsic RMs of fast radio bursts~\citep[e.g.][]{Pandhi+25} and to assess the detectability of possible polarization from the epoch of reionization~\citep{Li+21}.

\section{Summary}

Understanding the Galactic magnetic field is the first step toward understanding the origin and evolution of cosmic magnetism. However, there are still open questions on the Galactic magnetic field.  

An SKA-Mid polarization survey will produce an all-sky RM grid with a density of about 100~deg$^{-2}$, which is about two orders of magnitude greater than that currently being used to model the Galactic magnetic field. Together with single-dish observations, the survey will also produce total intensity, polarized intensity, and RM all-sky images of diffuse emission covering the scales from about $10\arcsec$ upward. The RM grid and the images of diffuse emission will lead to the most complete picture of the Galactic magnetic field, and thus advance our understanding of the origin and evolution of the magnetic field.     

\section*{Acknowledgments}
XS is supported by the National SKA Program of China (Grant No. 2022SKA0120101) and the National Natural Science Foundation of China (No. 12433006). AB acknowledges financial support from the INAF initiative ``IAF Astronomy Fellowships in Italy'' (grant name MEGASKAT).

\bibliographystyle{abbrvnat-maxbibnames4}
\bibliography{chapter} 

\begin{thebibliography}{108}
\providecommand{\natexlab}[1]{#1}
\providecommand{\url}[1]{\texttt{#1}}
\expandafter\ifx\csname urlstyle\endcsname\relax
  \providecommand{\doi}[1]{doi: #1}\else
  \providecommand{\doi}{doi: \begingroup \urlstyle{rm}\Url}\fi

\bibitem[{Anderson} et~al.(2021){Anderson}, {Heald}, {Eilek}, {Lenc}, {Gaensler}, {Rudnick}, {Van Eck}, {O'Sullivan}, {Stil}, {Chippendale}, {Riseley}, {Carretti}, {West}, {Farnes}, {Harvey-Smith}, {McClure-Griffiths}, {Bock}, {Bunton}, {Koribalski}, {Tremblay}, {Voronkov}, and {Warhurst}]{Anderson+21}
C.~S. {Anderson} et al.
\newblock \emph{\pasa}, 38:\penalty0 e020, Apr. 2021.
\newblock \doi{10.1017/pasa.2021.4}.

\bibitem[{Ball} et~al.(2023){Ball}, {Kothes}, {Rosolowsky}, {West}, {Becker}, {Filipovi{\'c}}, {Gaensler}, {Hopkins}, {Koribalski}, {Landecker}, {Leahy}, {Marvil}, {Sun}, {Bufano}, {Carretti}, {Ingallinera}, {Van Eck}, and {Willis}]{Ball+23}
B.~D. {Ball} et al.
\newblock \emph{\mnras}, 524\penalty0 (1):\penalty0 1396--1421, Sept. 2023.
\newblock \doi{10.1093/mnras/stad1953}.

\bibitem[{Ball} et~al.(2025){Ball}, {Kothes}, {Rosolowsky}, {Burger-Scheidlin}, {Filipovi{\'c}}, {Lazarevi{\'c}}, {Smeaton}, {Becker}, {Carretti}, {Gaensler}, {Hopkins}, {Leahy}, {Tahani}, {West}, {Anderson}, {Loru}, {Ma}, {McClure-Griffiths}, and {Micha{\l}owski}]{Ball+25}
B.~D. {Ball} et al.
\newblock \emph{\apj}, 988\penalty0 (1):\penalty0 75, July 2025.
\newblock \doi{10.3847/1538-4357/addc63}.

\bibitem[{Beck} and {Gaensler}(2004)]{Beck+04}
R.~{Beck} and B.~M. {Gaensler}.
\newblock \emph{\nar}, 48\penalty0 (11-12):\penalty0 1289--1304, Dec. 2004.
\newblock \doi{10.1016/j.newar.2004.09.013}.

\bibitem[{Beck} et~al.(2003){Beck}, {Shukurov}, {Sokoloff}, and {Wielebinski}]{Beck+03}
R.~{Beck}, A.~{Shukurov}, D.~{Sokoloff}, and R.~{Wielebinski}.
\newblock \emph{\aap}, 411:\penalty0 99--107, Nov. 2003.
\newblock \doi{10.1051/0004-6361:20031101}.

\bibitem[{Berger} et~al.(2025){Berger}, {Adebahr}, {Wright}, {Hildebrandt}, {Dettmar}, {Adams}, {D{\'e}nes}, {Hess}, {Morganti}, {Damstra}, {Kutkin}, {Loose}, {Mika}, {Oostrum}, {van Leeuwen}, and {Ziemke}]{Berger+25}
A.~{Berger} et al.
\newblock \emph{\aap}, 693:\penalty0 A202, Jan. 2025.
\newblock \doi{10.1051/0004-6361/202245733}.

\bibitem[{Boulanger} et~al.(2018){Boulanger}, {En{\ss}lin}, {Fletcher}, {Girichides}, {Hackstein}, {Haverkorn}, {H{\"o}randel}, {Jaffe}, {Jasche}, {Kachelrie{\ss}}, {Kotera}, {Pfrommer}, {Rachen}, {Rodrigues}, {Ruiz-Granados}, {Seta}, {Shukurov}, {Sigl}, {Steininger}, {Vacca}, {van der Velden}, {van Vliet}, and {Wang}]{Boulanger+18}
F.~{Boulanger} et al.
\newblock \emph{\jcap}, 2018\penalty0 (8):\penalty0 049, Aug. 2018.
\newblock \doi{10.1088/1475-7516/2018/08/049}.

\bibitem[{Bracco} et~al.(2024){Bracco}, {Padovani}, and {Galli}]{Bracco+24}
A.~{Bracco}, M.~{Padovani}, and D.~{Galli}.
\newblock \emph{\aap}, 686:\penalty0 A52, June 2024.
\newblock \doi{10.1051/0004-6361/202449625}.

\bibitem[{Brandenburg} and {Ntormousi}(2023)]{Brandenburg+23}
A.~{Brandenburg} and E.~{Ntormousi}.
\newblock \emph{\araa}, 61:\penalty0 561--606, Aug. 2023.
\newblock \doi{10.1146/annurev-astro-071221-052807}.

\bibitem[{Brentjens} and {de Bruyn}(2005)]{Brentjens+05}
M.~A. {Brentjens} and A.~G. {de Bruyn}.
\newblock \emph{\aap}, 441\penalty0 (3):\penalty0 1217--1228, Oct. 2005.
\newblock \doi{10.1051/0004-6361:20052990}.

\bibitem[{Brown} and {Taylor}(2001)]{Brown+01}
J.~C. {Brown} and A.~R. {Taylor}.
\newblock \emph{\apjl}, 563\penalty0 (1):\penalty0 L31--L34, Dec. 2001.
\newblock \doi{10.1086/338358}.

\bibitem[{Burn}(1966)]{Burn+66}
B.~J. {Burn}.
\newblock \emph{\mnras}, 133:\penalty0 67, Jan. 1966.
\newblock \doi{10.1093/mnras/133.1.67}.

\bibitem[{Carretti} et~al.(2019){Carretti}, {Haverkorn}, {Staveley-Smith}, {Bernardi}, {Gaensler}, {Kesteven}, {Poppi}, {Brown}, {Crocker}, {Purcell}, {Schnitzeler}, and {Sun}]{Carretti+19}
E.~{Carretti} et al.
\newblock \emph{\mnras}, 489\penalty0 (2):\penalty0 2330--2354, Oct. 2019.
\newblock \doi{10.1093/mnras/stz806}.

\bibitem[{Curtin} et~al.(2024){Curtin}, {Weisberg}, and {Rankin}]{Curtin+24}
A.~P. {Curtin}, J.~M. {Weisberg}, and J.~M. {Rankin}.
\newblock \emph{\apj}, 975\penalty0 (2):\penalty0 217, Nov. 2024.
\newblock \doi{10.3847/1538-4357/ad7b15}.

\bibitem[{Dickey} et~al.(2022){Dickey}, {West}, {Thomson}, {Landecker}, {Bracco}, {Carretti}, {Han}, {Hill}, {Ma}, {Mao}, {Ordog}, {Brown}, {Douglas}, {Erceg}, {Jeli{\'c}}, {Kothes}, and {Wolleben}]{Dickey+22}
J.~M. {Dickey} et al.
\newblock \emph{\apj}, 940\penalty0 (1):\penalty0 75, Nov. 2022.
\newblock \doi{10.3847/1538-4357/ac94ce}.

\bibitem[{Erceg} et~al.(2022){Erceg}, {Jeli{\'c}}, {Haverkorn}, {Bracco}, {Shimwell}, {Tasse}, {Dickey}, {Ceraj}, {Drabent}, {Hardcastle}, and {Turi{\'c}}]{Erceg+22}
A.~{Erceg} et al.
\newblock \emph{\aap}, 663:\penalty0 A7, July 2022.
\newblock \doi{10.1051/0004-6361/202142244}.

\bibitem[{Ferri{\`e}re} and {Terral}(2014)]{Ferriere+14}
K.~{Ferri{\`e}re} and P.~{Terral}.
\newblock \emph{\aap}, 561:\penalty0 A100, Jan. 2014.
\newblock \doi{10.1051/0004-6361/201322966}.

\bibitem[{Gaensler} et~al.(2001){Gaensler}, {Dickey}, {McClure-Griffiths}, {Green}, {Wieringa}, and {Haynes}]{Gaenlser+01}
B.~M. {Gaensler} et al.
\newblock \emph{\apj}, 549\penalty0 (2):\penalty0 959--978, Mar. 2001.
\newblock \doi{10.1086/319468}.

\bibitem[{Gaensler} et~al.(2011){Gaensler}, {Haverkorn}, {Burkhart}, {Newton-McGee}, {Ekers}, {Lazarian}, {McClure-Griffiths}, {Robishaw}, {Dickey}, and {Green}]{Gaensler+11}
B.~M. {Gaensler} et al.
\newblock \emph{\nat}, 478\penalty0 (7368):\penalty0 214--217, Oct. 2011.
\newblock \doi{10.1038/nature10446}.

\bibitem[{Gaensler} et~al.(2025){Gaensler}, {Heald}, {McClure-Griffiths}, {Anderson}, {Van Eck}, {West}, {Thomson}, {Leahy}, {Rudnick}, {Ma}, {Akahori}, {G{\"u}rkan}, {Landecker}, {Mao}, {O'Sullivan}, {Raja}, {Sun}, {Vernstrom}, {Baidoo}, {Carretti}, {Taylor}, {Willis}, {Osinga}, {Livingston}, {Alexander}, {Alonso-L{\'o}pez}, {Amaral}, {An}, {Bracco}, {Bradbury}, {Br{\"u}ggen}, {Eswaraiah}, {En{\ss}lin}, {Galvin}, {Haverkorn}, {Hopkins}, {Hutschenreuter}, {Ideguchi}, {Jaswanth}, {Jung}, {Kaczmarek}, {Kothes}, {Lazarevi{\'c}}, {Leahy}, {Loi}, {Marvil}, {Norris}, {Pandhi}, {Price}, {Riseley}, {Ryder}, {Seta}, {Shaw}, {Shen}, {Sobey}, {Stil}, {Stuardi}, {Upasana}, {Vanderwoude}, and {Velovi{\'c}}]{Gaensler+25}
B.~M. {Gaensler} et al.
\newblock \emph{\pasa}, 42:\penalty0 e091, June 2025.
\newblock \doi{10.1017/pasa.2025.10031}.

\bibitem[{Han}(2017)]{Han+17}
J.~L. {Han}.
\newblock \emph{\araa}, 55\penalty0 (1):\penalty0 111--157, Aug. 2017.
\newblock \doi{10.1146/annurev-astro-091916-055221}.

\bibitem[{Han} et~al.(1999){Han}, {Manchester}, and {Qiao}]{Han+99}
J.~L. {Han}, R.~N. {Manchester}, and G.~J. {Qiao}.
\newblock \emph{\mnras}, 306\penalty0 (2):\penalty0 371--380, June 1999.
\newblock \doi{10.1046/j.1365-8711.1999.02544.x}.

\bibitem[{Han} et~al.(2018){Han}, {Manchester}, {van Straten}, and {Demorest}]{Han+18}
J.~L. {Han}, R.~N. {Manchester}, W.~{van Straten}, and P.~{Demorest}.
\newblock \emph{\apjs}, 234\penalty0 (1):\penalty0 11, Jan. 2018.
\newblock \doi{10.3847/1538-4365/aa9c45}.

\bibitem[{Haverkorn}(2015)]{Haverkorn+15b}
M.~{Haverkorn}.
\newblock In A.~{Lazarian}, E.~M. {de Gouveia Dal Pino}, and C.~{Melioli}, editors, \emph{Magnetic Fields in Diffuse Media}, volume 407 of \emph{Astrophysics and Space Science Library}, page 483, Jan. 2015.
\newblock \doi{10.1007/978-3-662-44625-6_17}.

\bibitem[{Haverkorn} et~al.(2000){Haverkorn}, {Katgert}, and {de Bruyn}]{Haverkorn+00}
M.~{Haverkorn}, P.~{Katgert}, and A.~G. {de Bruyn}.
\newblock \emph{\aap}, 356:\penalty0 L13--L16, Apr. 2000.
\newblock \doi{10.48550/arXiv.astro-ph/0003260}.

\bibitem[{Haverkorn} et~al.(2008){Haverkorn}, {Brown}, {Gaensler}, and {McClure-Griffiths}]{Haverkorn+08}
M.~{Haverkorn}, J.~C. {Brown}, B.~M. {Gaensler}, and N.~M. {McClure-Griffiths}.
\newblock \emph{\apj}, 680\penalty0 (1):\penalty0 362--370, June 2008.
\newblock \doi{10.1086/587165}.

\bibitem[{Haverkorn} et~al.(2015){Haverkorn}, {Akahori}, {Carretti}, {Ferri{\`e}re}, {Frick}, {Gaensler}, {Heald}, {Johnston-Hollitt}, {Jones}, {Landecker}, {Mao}, {Noutsos}, {Oppermann}, {Reich}, {Robishaw}, {Scaife}, {Schnitzeler}, {Stepanov}, {Sun}, and {Taylor}]{Haverkorn+15}
M.~{Haverkorn} et al.
\newblock In \emph{Advancing Astrophysics with the Square Kilometre Array (AASKA14)}, page~96, Apr. 2015.
\newblock \doi{10.22323/1.215.0096}.

\bibitem[{Heald} et~al.(2020){Heald}, {Mao}, {Vacca}, {Akahori}, {Damas-Segovia}, {Gaensler}, {Hoeft}, {Agudo}, {Basu}, {Beck}, {Birkinshaw}, {Bonafede}, {Bourke}, {Bracco}, {Carretti}, {Feretti}, {Girart}, {Govoni}, {Green}, {Han}, {Haverkorn}, {Horellou}, {Johnston-Hollitt}, {Kothes}, {Landecker}, {Nikiel-Wroczy{\'n}ski}, {O'Sullivan}, {Padovani}, {Poidevin}, {Pratley}, {Regis}, {Riseley}, {Robishaw}, {Rudnick}, {Sobey}, {Stil}, {Sun}, {Sur}, {Taylor}, {Thomson}, {Van Eck}, {Vazza}, {West}, and {the SKA Magnetism Science Working Group}]{Heald+20}
G.~{Heald} et al.
\newblock \emph{Galaxies}, 8\penalty0 (3):\penalty0 53, July 2020.
\newblock \doi{10.3390/galaxies8030053}.

\bibitem[{Hoang} and {Lazarian}(2016)]{HoangLaz2016}
T.~{Hoang} and A.~{Lazarian}.
\newblock \emph{\apj}, 831\penalty0 (2):\penalty0 159, Nov. 2016.
\newblock \doi{10.3847/0004-637X/831/2/159}.

\bibitem[{Hoang} and {Truong}(2024)]{HoangTruong2024}
T.~{Hoang} and B.~{Truong}.
\newblock \emph{\apj}, 965\penalty0 (2):\penalty0 183, Apr. 2024.
\newblock \doi{10.3847/1538-4357/ad2a56}.

\bibitem[{Hopkins} et~al.(2025){Hopkins}, {Kapinska}, {Marvil}, {Vernstrom}, {Collier}, {Norris}, {Gordon}, {Duchesne}, {Rudnick}, {Gupta}, {Carretti}, {Anderson}, {Dai}, {G{\"u}rkan}, {Parkinson}, {Prandoni}, {Riggi}, {Shekhar Saraf}, {Ma}, {Filipovi{\'c}}, {Umana}, {Bahr-Kalus}, {Koribalski}, {Lenc}, {Ingallinera}, {Afonso}, {Ahmad}, {Ahmed}, {Alexander}, {Andernach}, {Asorey}, {Battisti}, {Bilicki}, {Botteon}, {Brown}, {Br{\"u}ggen}, {Cowley}, {Dage}, {Hale}, {Hardcastle}, {Kothes}, {Lazarevi{\'c}}, {Lin}, {Luken}, {Moss}, {Prathap}, {ur Rahman}, {Reiprich}, {Riseley}, {Salvato}, {Seymour}, {Shabala}, {Smith}, {Vaccari}, {van Loon}, {Wong}, {Zainal Alsaberi}, {Asher}, {Ball}, {Barbosa}, {Biava}, {Bradley}, {Carvajal}, {Crawford}, {Galvin}, {Huynh}, {Leahy}, {Matute}, {Moss}, {Pappalardo}, {Smeaton}, {Velovi{\'c}}, and {Zafar}]{Hopkins+25}
A.~{Hopkins} et al.
\newblock \emph{\pasa}, 42:\penalty0 e071, May 2025.
\newblock \doi{10.1017/pasa.2025.10042}.

\bibitem[{Hutschenreuter} et~al.(2022){Hutschenreuter}, {Anderson}, {Betti}, {Bower}, {Brown}, {Br{\"u}ggen}, {Carretti}, {Clarke}, {Clegg}, {Costa}, {Croft}, {Van Eck}, {Gaensler}, {de Gasperin}, {Haverkorn}, {Heald}, {Hull}, {Inoue}, {Johnston-Hollitt}, {Kaczmarek}, {Law}, {Ma}, {MacMahon}, {Mao}, {Riseley}, {Roy}, {Shanahan}, {Shimwell}, {Stil}, {Sobey}, {O'Sullivan}, {Tasse}, {Vacca}, {Vernstrom}, {Williams}, {Wright}, and {En{\ss}lin}]{Hutschenreuter+22}
S.~{Hutschenreuter} et al.
\newblock \emph{\aap}, 657:\penalty0 A43, Jan. 2022.
\newblock \doi{10.1051/0004-6361/202140486}.

\bibitem[{Jaffe}(2019)]{Jaffe+19}
T.~R. {Jaffe}.
\newblock \emph{Galaxies}, 7\penalty0 (2):\penalty0 52, Apr. 2019.
\newblock \doi{10.3390/galaxies7020052}.

\bibitem[{Jaffe} et~al.(2010){Jaffe}, {Leahy}, {Banday}, {Leach}, {Lowe}, and {Wilkinson}]{Jaffe+10}
T.~R. {Jaffe} et al.
\newblock \emph{\mnras}, 401\penalty0 (2):\penalty0 1013--1028, Jan. 2010.
\newblock \doi{10.1111/j.1365-2966.2009.15745.x}.

\bibitem[{Jansson} and {Farrar}(2012)]{Jansson+12}
R.~{Jansson} and G.~R. {Farrar}.
\newblock \emph{\apjl}, 761\penalty0 (1):\penalty0 L11, Dec. 2012.
\newblock \doi{10.1088/2041-8205/761/1/L11}.

\bibitem[{Jiao} et~al.(2022){Jiao}, {Lin}, {Shui}, {Wu}, {Ren}, and {Li}]{Jiao+22}
S.~{Jiao} et al.
\newblock \emph{Science China Physics, Mechanics, and Astronomy}, 65\penalty0 (9):\penalty0 299511, Sept. 2022.
\newblock \doi{10.1007/s11433-021-1902-3}.

\bibitem[{Johnston-Hollitt} et~al.(2015){Johnston-Hollitt}, {Govoni}, {Beck}, {Dehghan}, {Pratley}, {Akahori}, {Heald}, {Agudo}, {Bonafede}, {Carretti}, {Clarke}, {Colafrancesco}, {Ensslin}, {Feretti}, {Gaensler}, {Haverkorn}, {Mao}, {Oppermann}, {Rudnick}, {Scaife}, {Schnitzeler}, {Stil}, {Taylor}, and {Vacca}]{Johnston+15}
M.~{Johnston-Hollitt} et al.
\newblock In \emph{Advancing Astrophysics with the Square Kilometre Array (AASKA14)}, page~92, Apr. 2015.
\newblock \doi{10.22323/1.215.0092}.

\bibitem[{Khadir} et~al.(2024){Khadir}, {Pandhi}, {Hutschenreuter}, {Gaensler}, {Vanderwoude}, {West}, and {O'Sullivan}]{Khadir+24}
A.~{Khadir} et al.
\newblock \emph{\apj}, 977\penalty0 (2):\penalty0 276, Dec. 2024.
\newblock \doi{10.3847/1538-4357/ad8ddf}.

\bibitem[{Koda} et~al.(2019){Koda}, {Teuben}, {Sawada}, {Plunkett}, and {Fomalont}]{Koda+19}
J.~{Koda} et al.
\newblock \emph{\pasp}, 131\penalty0 (999):\penalty0 054505, May 2019.
\newblock \doi{10.1088/1538-3873/ab047e}.

\bibitem[{Krause} et~al.(2020){Krause}, {Irwin}, {Schmidt}, {Stein}, {Miskolczi}, {Carolina Mora-Partiarroyo}, {Wiegert}, {Beck}, {Stil}, {Heald}, {Li}, {Damas-Segovia}, {Vargas}, {Rand}, {West}, {Walterbos}, {Dettmar}, {English}, and {Woodfinden}]{Krause+20}
M.~{Krause} et al.
\newblock \emph{\aap}, 639:\penalty0 A112, July 2020.
\newblock \doi{10.1051/0004-6361/202037780}.

\bibitem[{Lacy} et~al.(2020){Lacy}, {Baum}, {Chandler}, {Chatterjee}, {Clarke}, {Deustua}, {English}, {Farnes}, {Gaensler}, {Gugliucci}, {Hallinan}, {Kent}, {Kimball}, {Law}, {Lazio}, {Marvil}, {Mao}, {Medlin}, {Mooley}, {Murphy}, {Myers}, {Osten}, {Richards}, {Rosolowsky}, {Rudnick}, {Schinzel}, {Sivakoff}, {Sjouwerman}, {Taylor}, {White}, {Wrobel}, {Andernach}, {Beasley}, {Berger}, {Bhatnager}, {Birkinshaw}, {Bower}, {Brandt}, {Brown}, {Burke-Spolaor}, {Butler}, {Comerford}, {Demorest}, {Fu}, {Giacintucci}, {Golap}, {G{\"u}th}, {Hales}, {Hiriart}, {Hodge}, {Horesh}, {Ivezi{\'c}}, {Jarvis}, {Kamble}, {Kassim}, {Liu}, {Loinard}, {Lyons}, {Masters}, {Mezcua}, {Moellenbrock}, {Mroczkowski}, {Nyland}, {O'Dea}, {O'Sullivan}, {Peters}, {Radford}, {Rao}, {Robnett}, {Salcido}, {Shen}, {Sobotka}, {Witz}, {Vaccari}, {van Weeren}, {Vargas}, {Williams}, and {Yoon}]{Lacy+20}
M.~{Lacy} et al.
\newblock \emph{\pasp}, 132\penalty0 (1009):\penalty0 035001, Mar. 2020.
\newblock \doi{10.1088/1538-3873/ab63eb}.

\bibitem[{Lazarian} et~al.(2024){Lazarian}, {Yuen}, and {Pogosyan}]{Lazarian+24}
A.~{Lazarian}, K.~H. {Yuen}, and D.~{Pogosyan}.
\newblock \emph{\apj}, 974\penalty0 (2):\penalty0 237, Oct. 2024.
\newblock \doi{10.3847/1538-4357/ad6d62}.

\bibitem[{Li} et~al.(2021){Li}, {Tan}, and {Mao}]{Li+21}
B.~{Li}, J.~{Tan}, and Y.~{Mao}.
\newblock \emph{\apj}, 918\penalty0 (1):\penalty0 14, Sept. 2021.
\newblock \doi{10.3847/1538-4357/ac09e4}.

\bibitem[{Li} et~al.(2018){Li}, {Wang}, {Qian}, {Krco}, {Jiang}, {Yue}, {Jin}, {Zhu}, {Pan}, {Nan}, and {Dunning}]{Li+18}
D.~{Li} et al.
\newblock \emph{IEEE Microwave Magazine}, 19\penalty0 (3):\penalty0 112--119, Apr. 2018.
\newblock \doi{10.1109/MMM.2018.2802178}.

\bibitem[{Loi} et~al.(2025){Loi}, {Serra}, {Murgia}, {Govoni}, {Vacca}, {Maccagni}, {Kleiner}, and {Kamphuis}]{Loi+25}
F.~{Loi} et al.
\newblock \emph{\aap}, 694:\penalty0 A125, Feb. 2025.
\newblock \doi{10.1051/0004-6361/202451711}.

\bibitem[{Ma} et~al.(2019){Ma}, {Mao}, {Stil}, {Basu}, {West}, {Heiles}, {Hill}, and {Betti}]{Ma+19a}
Y.~K. {Ma} et al.
\newblock \emph{\mnras}, 487\penalty0 (3):\penalty0 3432--3453, Aug. 2019.
\newblock \doi{10.1093/mnras/stz1325}.

\bibitem[{Ma} et~al.(2020){Ma}, {Mao}, {Ordog}, and {Brown}]{Ma+20}
Y.~K. {Ma}, S.~A. {Mao}, A.~{Ordog}, and J.~C. {Brown}.
\newblock \emph{\mnras}, 497\penalty0 (3):\penalty0 3097--3117, Sept. 2020.
\newblock \doi{10.1093/mnras/staa2105}.

\bibitem[{Ma} et~al.(2025){Ma}, {Seta}, {McClure-Griffiths}, {Van Eck}, {Mao}, {Ordog}, {Brown}, {Kovacs}, {Akahori}, {Kurahara}, {Oberhelman}, and {Anderson}]{Ma+25}
Y.~K. {Ma} et al.
\newblock \emph{\mnras}, 541\penalty0 (1):\penalty0 306--336, July 2025.
\newblock \doi{10.1093/mnras/staf1000}.

\bibitem[Ma et~al.(2026)Ma, author2, author3, author4, and author5]{Ma01.2026.SKA}
Y.~K. Ma et al.
\newblock In \emph{Advancing Astrophysics with the SKA -- II (AASKAII)}. 2026.
\newblock arXiv search: Report number AASKAII/Ma01.

\bibitem[{Maconi} et~al.(2025){Maconi}, {Reissl}, {Soler}, {Girichidis}, {Klessen}, {Bracco}, and {Hutschenreuter}]{Maconi+25}
E.~{Maconi} et al.
\newblock \emph{\aap}, 698:\penalty0 A84, June 2025.
\newblock \doi{10.1051/0004-6361/202451477}.

\bibitem[{Mao} et~al.(2010){Mao}, {Gaensler}, {Haverkorn}, {Zweibel}, {Madsen}, {McClure-Griffiths}, {Shukurov}, and {Kronberg}]{Mao+10}
S.~A. {Mao} et al.
\newblock \emph{\apj}, 714\penalty0 (2):\penalty0 1170--1186, May 2010.
\newblock \doi{10.1088/0004-637X/714/2/1170}.

\bibitem[{Mao} et~al.(2012){Mao}, {McClure-Griffiths}, {Gaensler}, {Brown}, {van Eck}, {Haverkorn}, {Kronberg}, {Stil}, {Shukurov}, and {Taylor}]{Mao+12}
S.~A. {Mao} et al.
\newblock \emph{\apj}, 755\penalty0 (1):\penalty0 21, Aug. 2012.
\newblock \doi{10.1088/0004-637X/755/1/21}.

\bibitem[Mao et~al.(2026)Mao, author2, author3, author4, and author5]{Mao01.2026.SKA}
S.~A. Mao et al.
\newblock In \emph{Advancing Astrophysics with the SKA -- II (AASKAII)}. 2026.
\newblock arXiv search: Report number AASKAII/Mao01.

\bibitem[{Marinacci} and {Vogelsberger}(2016)]{Marinacci+16}
F.~{Marinacci} and M.~{Vogelsberger}.
\newblock \emph{\mnras}, 456\penalty0 (1):\penalty0 L69--L73, Feb. 2016.
\newblock \doi{10.1093/mnrasl/slv176}.

\bibitem[{Minter} and {Spangler}(1996)]{Minter+96}
A.~H. {Minter} and S.~R. {Spangler}.
\newblock \emph{\apj}, 458:\penalty0 194, Feb. 1996.
\newblock \doi{10.1086/176803}.

\bibitem[{Ocker} and {Cordes}(2026)]{Ocker+26}
S.~K. {Ocker} and J.~M. {Cordes}.
\newblock \emph{arXiv e-prints}, art. arXiv:2602.11838, Feb. 2026.
\newblock \doi{10.48550/arXiv.2602.11838}.

\bibitem[{Ordog} et~al.(2017){Ordog}, {Brown}, {Kothes}, and {Landecker}]{Ordog+17}
A.~{Ordog}, J.~C. {Brown}, R.~{Kothes}, and T.~L. {Landecker}.
\newblock \emph{\aap}, 603:\penalty0 A15, July 2017.
\newblock \doi{10.1051/0004-6361/201730740}.

\bibitem[{Ordog} et~al.(2025){Ordog}, {Brown}, {Landecker}, {Hill}, {Kothes}, {West}, {Dickey}, {Haverkorn}, {Carretti}, {Thomson}, {Bracco}, {Del Rizzo}, {Ransom}, and {Reid}]{Ordog+25}
A.~{Ordog} et al.
\newblock \emph{\aj}, 169\penalty0 (6):\penalty0 312, June 2025.
\newblock \doi{10.3847/1538-3881/adc929}.

\bibitem[{Ordog} et~al.(2026){Ordog}, {Booth}, {Landecker}, {Carretti}, {Hill}, {Brown}, {Davydov}, {Caffarello}, {Galler}, {Flygare}, {West}, {Willis}, {Tahani}, {Hovey}, {Lagoy}, {Harrison}, {Smith}, {Baard}, {Messing}, {Del Rizzo}, {Robert}, {Robishaw}, {Dickey}, {Morgan}, {Kennedy}, {Haverkorn}, {Bracco}, and {Conway}]{Ordog+26}
A.~{Ordog} et al.
\newblock \emph{\apjs}, 282\penalty0 (2):\penalty0 53, Feb. 2026.
\newblock \doi{10.3847/1538-4365/ae2471}.

\bibitem[{O'Sullivan} et~al.(2023){O'Sullivan}, {Shimwell}, {Hardcastle}, {Tasse}, {Heald}, {Carretti}, {Br{\"u}ggen}, {Vacca}, {Sobey}, {Van Eck}, {Horellou}, {Beck}, {Bilicki}, {Bourke}, {Botteon}, {Croston}, {Drabent}, {Duncan}, {Heesen}, {Ideguchi}, {Kirwan}, {Lawlor}, {Mingo}, {Nikiel-Wroczy{\'n}ski}, {Piotrowska}, {Scaife}, and {van Weeren}]{OSullivan+23}
S.~P. {O'Sullivan} et al.
\newblock \emph{\mnras}, 519\penalty0 (4):\penalty0 5723--5742, Mar. 2023.
\newblock \doi{10.1093/mnras/stac3820}.

\bibitem[{Oswald} et~al.(2025){Oswald}, {Weltevrede}, {Posselt}, {Johnston}, {Karastergiou}, and {Lower}]{Oswald+25}
L.~S. {Oswald} et al.
\newblock \emph{\mnras}, 540\penalty0 (3):\penalty0 2112--2130, July 2025.
\newblock \doi{10.1093/mnras/staf645}.

\bibitem[{Pandhi} et~al.(2025){Pandhi}, {Gaensler}, {Pleunis}, {Hutschenreuter}, {Law}, {Mckinven}, {O'Sullivan}, {Petroff}, and {Vernstrom}]{Pandhi+25}
A.~{Pandhi} et al.
\newblock \emph{\apj}, 982\penalty0 (2):\penalty0 146, Apr. 2025.
\newblock \doi{10.3847/1538-4357/adb8e3}.

\bibitem[{Pattle} et~al.(2023){Pattle}, {Fissel}, {Tahani}, {Liu}, and {Ntormousi}]{Pattle+23}
K.~{Pattle} et al.
\newblock In S.~{Inutsuka} et al., editors, \emph{Protostars and Planets VII}, volume 534 of \emph{Astronomical Society of the Pacific Conference Series}, page 193, July 2023.
\newblock \doi{10.48550/arXiv.2203.11179}.

\bibitem[{Pelgrims} et~al.(2021){Pelgrims}, {Mac{\'\i}as-P{\'e}rez}, and {Ruppin}]{Pelgrims+21}
V.~{Pelgrims}, J.~F. {Mac{\'\i}as-P{\'e}rez}, and F.~{Ruppin}.
\newblock \emph{\aap}, 652:\penalty0 A130, Aug. 2021.
\newblock \doi{10.1051/0004-6361/201833962}.

\bibitem[{Planck Collaboration} et~al.(2016){Planck Collaboration}, {Adam}, {Ade}, {Alves}, {Ashdown}, {Aumont}, {Baccigalupi}, {Banday}, {Barreiro}, {Bartolo}, {Battaner}, {Benabed}, {Benoit-L{\'e}vy}, {Bernard}, {Bersanelli}, {Bielewicz}, {Bonavera}, {Bond}, {Borrill}, {Bouchet}, {Boulanger}, {Bucher}, {Burigana}, {Butler}, {Calabrese}, {Cardoso}, {Catalano}, {Chiang}, {Christensen}, {Colombo}, {Combet}, {Couchot}, {Crill}, {Curto}, {Cuttaia}, {Danese}, {Davis}, {de Bernardis}, {de Rosa}, {de Zotti}, {Delabrouille}, {Dickinson}, {Diego}, {Dolag}, {Dor{\'e}}, {Ducout}, {Dupac}, {Elsner}, {En{\ss}lin}, {Eriksen}, {Ferri{\`e}re}, {Finelli}, {Forni}, {Frailis}, {Fraisse}, {Franceschi}, {Galeotta}, {Ganga}, {Ghosh}, {Giard}, {Gjerl{\o}w}, {Gonz{\'a}lez-Nuevo}, {G{\'o}rski}, {Gregorio}, {Gruppuso}, {Gudmundsson}, {Hansen}, {Harrison}, {Hern{\'a}ndez-Monteagudo}, {Herranz}, {Hildebrandt}, {Hobson}, {Hornstrup}, {Hurier}, {Jaffe}, {Jaffe}, {Jones}, {Juvela}, {Keih{\"a}nen}, {Keskitalo}, {Kisner}, {Knoche}, {Kunz},
  {Kurki-Suonio}, {Lamarre}, {Lasenby}, {Lattanzi}, {Lawrence}, {Leahy}, {Leonardi}, {Levrier}, {Liguori}, {Lilje}, {Linden-V{\o}rnle}, {L{\'o}pez-Caniego}, {Lubin}, {Mac{\'\i}as-P{\'e}rez}, {Maggio}, {Maino}, {Mandolesi}, {Mangilli}, {Maris}, {Martin}, {Mart{\'\i}nez-Gonz{\'a}lez}, {Masi}, {Matarrese}, {Melchiorri}, {Mennella}, {Migliaccio}, {Miville-Desch{\^e}nes}, {Moneti}, {Montier}, {Morgante}, {Munshi}, {Murphy}, {Naselsky}, {Nati}, {Natoli}, {N{\o}rgaard-Nielsen}, {Oppermann}, {Orlando}, {Pagano}, {Pajot}, {Paladini}, {Paoletti}, {Pasian}, {Perotto}, {Pettorino}, {Piacentini}, {Piat}, {Pierpaoli}, {Plaszczynski}, {Pointecouteau}, {Polenta}, {Ponthieu}, {Pratt}, {Prunet}, {Puget}, {Rachen}, {Reinecke}, {Remazeilles}, {Renault}, {Renzi}, {Ristorcelli}, {Rocha}, {Rossetti}, {Roudier}, {Rubi{\~n}o-Mart{\'\i}n}, {Rusholme}, {Sandri}, {Santos}, {Savelainen}, {Scott}, {Spencer}, {Stolyarov}, {Stompor}, {Strong}, {Sudiwala}, {Sunyaev}, {Suur-Uski}, {Sygnet}, {Tauber}, {Terenzi}, {Toffolatti}, {Tomasi},
  {Tristram}, {Tucci}, {Valenziano}, {Valiviita}, {Van Tent}, {Vielva}, {Villa}, {Wade}, {Wandelt}, {Wehus}, {Yvon}, {Zacchei}, and {Zonca}]{Adam+16}
{Planck Collaboration} et al.
\newblock \emph{\aap}, 596:\penalty0 A103, Dec. 2016.
\newblock \doi{10.1051/0004-6361/201528033}.

\bibitem[{Plunkett} et~al.(2023){Plunkett}, {Hacar}, {Moser-Fischer}, {Petry}, {Teuben}, {Pingel}, {Kunneriath}, {Takagi}, {Miyamoto}, {Moravec}, {Suri}, {Hess}, {Hoffman}, and {Mason}]{Plunkett+23}
A.~{Plunkett} et al.
\newblock \emph{\pasp}, 135\penalty0 (1045):\penalty0 034501, Mar. 2023.
\newblock \doi{10.1088/1538-3873/acb9bd}.

\bibitem[{Prandoni} and {Seymour}(2015)]{Prandoni+15}
I.~{Prandoni} and N.~{Seymour}.
\newblock In \emph{Advancing Astrophysics with the Square Kilometre Array (AASKA14)}, page~67, Apr. 2015.
\newblock \doi{10.22323/1.215.0067}.

\bibitem[{Purcell} et~al.(2020){Purcell}, {Van Eck}, {West}, {Sun}, and {Gaensler}]{Purcell+20}
C.~R. {Purcell} et al.
\newblock {RM-Tools: Rotation measure (RM) synthesis and Stokes QU-fitting}.
\newblock Astrophysics Source Code Library, record ascl:2005.003, May 2020.

\bibitem[{Ranchod} et~al.(2024){Ranchod}, {Mao}, {Deane}, {Sridhar}, {Damas-Segovia}, {Livingston}, and {Ma}]{Ranchod+24}
S.~{Ranchod} et al.
\newblock \emph{\aap}, 686:\penalty0 A104, June 2024.
\newblock \doi{10.1051/0004-6361/202348993}.

\bibitem[{Rau} et~al.(2019){Rau}, {Naik}, and {Braun}]{Rau+19}
U.~{Rau}, N.~{Naik}, and T.~{Braun}.
\newblock \emph{\aj}, 158\penalty0 (1):\penalty0 3, July 2019.
\newblock \doi{10.3847/1538-3881/ab1aa7}.

\bibitem[{Riseley} et~al.(2020){Riseley}, {Galvin}, {Sobey}, {Vernstrom}, {White}, {Zhang}, {Gaensler}, {Heald}, {Anderson}, {Franzen}, {Hancock}, {Hurley-Walker}, {Lenc}, and {Van Eck}]{Riseley+20}
C.~J. {Riseley} et al.
\newblock \emph{\pasa}, 37:\penalty0 e029, July 2020.
\newblock \doi{10.1017/pasa.2020.20}.

\bibitem[{Rudnick} and {Owen}(2014)]{Rudnick+14}
L.~{Rudnick} and F.~N. {Owen}.
\newblock \emph{\apj}, 785\penalty0 (1):\penalty0 45, Apr. 2014.
\newblock \doi{10.1088/0004-637X/785/1/45}.

\bibitem[{Schnitzeler}(2010)]{Schnitzeler+10}
D.~H.~F.~M. {Schnitzeler}.
\newblock \emph{\mnras}, 409\penalty0 (1):\penalty0 L99--L103, Nov. 2010.
\newblock \doi{10.1111/j.1745-3933.2010.00957.x}.

\bibitem[{Schnitzeler} et~al.(2019){Schnitzeler}, {Carretti}, {Wieringa}, {Gaensler}, {Haverkorn}, and {Poppi}]{Schnitzeler+19}
D.~H.~F.~M. {Schnitzeler} et al.
\newblock \emph{\mnras}, 485\penalty0 (1):\penalty0 1293--1309, May 2019.
\newblock \doi{10.1093/mnras/stz092}.

\bibitem[{Seta} and {Federrath}(2020)]{Seta+20}
A.~{Seta} and C.~{Federrath}.
\newblock \emph{\mnras}, 499\penalty0 (2):\penalty0 2076--2086, Dec. 2020.
\newblock \doi{10.1093/mnras/staa2978}.

\bibitem[{Seta} and {Federrath}(2021)]{Seta+21}
A.~{Seta} and C.~{Federrath}.
\newblock \emph{\mnras}, 502\penalty0 (2):\penalty0 2220--2237, Apr. 2021.
\newblock \doi{10.1093/mnras/stab128}.

\bibitem[{Shukurov} and {Subramanian}(2022)]{Shukurov+22}
A.~M. {Shukurov} and K.~{Subramanian}.
\newblock \emph{{Astrophysical Magnetic Fields: From Galaxies to the Early Universe}}.
\newblock 2022.
\newblock \doi{10.1017/9781139046657}.

\bibitem[{Sokoloff} et~al.(1998){Sokoloff}, {Bykov}, {Shukurov}, {Berkhuijsen}, {Beck}, and {Poezd}]{Sokoloff+98}
D.~D. {Sokoloff} et al.
\newblock \emph{\mnras}, 299\penalty0 (1):\penalty0 189--206, Aug. 1998.
\newblock \doi{10.1046/j.1365-8711.1998.01782.x}.

\bibitem[{Stil} et~al.(2011){Stil}, {Taylor}, and {Sunstrum}]{Stil+11}
J.~M. {Stil}, A.~R. {Taylor}, and C.~{Sunstrum}.
\newblock \emph{\apj}, 726\penalty0 (1):\penalty0 4, Jan. 2011.
\newblock \doi{10.1088/0004-637X/726/1/4}.

\bibitem[{Subramanian}(2016)]{Subramanian+16}
K.~{Subramanian}.
\newblock \emph{Reports on Progress in Physics}, 79\penalty0 (7):\penalty0 076901, July 2016.
\newblock \doi{10.1088/0034-4885/79/7/076901}.

\bibitem[{Sun} et~al.(2025){Sun}, {Haverkorn}, {Carretti}, {Landecker}, {Gaensler}, {Poppi}, {Staveley-Smith}, {Gao}, and {Han}]{Sun+25}
X.~{Sun} et al.
\newblock \emph{\aap}, 694:\penalty0 A169, Feb. 2025.
\newblock \doi{10.1051/0004-6361/202453326}.

\bibitem[{Sun} and {Reich}(2009)]{Sun+09}
X.~H. {Sun} and W.~{Reich}.
\newblock \emph{\aap}, 507\penalty0 (2):\penalty0 1087--1105, Nov. 2009.
\newblock \doi{10.1051/0004-6361/200912539}.

\bibitem[{Sun} et~al.(2007){Sun}, {Han}, {Reich}, {Reich}, {Shi}, {Wielebinski}, and {F{\"u}rst}]{Sun+07}
X.~H. {Sun} et al.
\newblock \emph{\aap}, 463\penalty0 (3):\penalty0 993--1007, Mar. 2007.
\newblock \doi{10.1051/0004-6361:20066001}.

\bibitem[{Sun} et~al.(2008){Sun}, {Reich}, {Waelkens}, and {En{\ss}lin}]{Sun+08}
X.~H. {Sun}, W.~{Reich}, A.~{Waelkens}, and T.~A. {En{\ss}lin}.
\newblock \emph{\aap}, 477\penalty0 (2):\penalty0 573--592, Jan. 2008.
\newblock \doi{10.1051/0004-6361:20078671}.

\bibitem[{Sun} et~al.(2015{\natexlab{a}}){Sun}, {Landecker}, {Gaensler}, {Carretti}, {Reich}, {Leahy}, {McClure-Griffiths}, {Crocker}, {Wolleben}, {Haverkorn}, {Douglas}, and {Gray}]{Sun+15}
X.~H. {Sun} et al.
\newblock \emph{\apj}, 811\penalty0 (1):\penalty0 40, Sept. 2015{\natexlab{a}}.
\newblock \doi{10.1088/0004-637X/811/1/40}.

\bibitem[{Sun} et~al.(2015{\natexlab{b}}){Sun}, {Rudnick}, {Akahori}, {Anderson}, {Bell}, {Bray}, {Farnes}, {Ideguchi}, {Kumazaki}, {O'Brien}, {O'Sullivan}, {Scaife}, {Stepanov}, {Stil}, {Takahashi}, {van Weeren}, and {Wolleben}]{Sun+15b}
X.~H. {Sun} et al.
\newblock \emph{\aj}, 149\penalty0 (2):\penalty0 60, Feb. 2015{\natexlab{b}}.
\newblock \doi{10.1088/0004-6256/149/2/60}.

\bibitem[{Sun} et~al.(2022){Sun}, {Gao}, {Reich}, {Jiang}, {Li}, {Yan}, and {Li}]{Sun+22}
X.-H. {Sun} et al.
\newblock \emph{Research in Astronomy and Astrophysics}, 22\penalty0 (12):\penalty0 125011, Dec. 2022.
\newblock \doi{10.1088/1674-4527/ac9d27}.

\bibitem[Tahani et~al.(2026)Tahani, author2, author3, author4, and author5]{Tahani01.2026.SKA}
M.~Tahani et al.
\newblock In \emph{Advancing Astrophysics with the SKA -- II (AASKAII)}. 2026.
\newblock arXiv search: Report number AASKAII/Tahani01.

\bibitem[{Taylor} et~al.(2009){Taylor}, {Stil}, and {Sunstrum}]{Taylor+09}
A.~R. {Taylor}, J.~M. {Stil}, and C.~{Sunstrum}.
\newblock \emph{\apj}, 702\penalty0 (2):\penalty0 1230--1236, Sept. 2009.
\newblock \doi{10.1088/0004-637X/702/2/1230}.

\bibitem[{Taylor} et~al.(2024){Taylor}, {Sekhar}, {Heino}, {Scaife}, {Stil}, {Bowles}, {Jarvis}, {Heywood}, and {Collier}]{Taylor+24}
A.~R. {Taylor} et al.
\newblock \emph{\mnras}, 528\penalty0 (2):\penalty0 2511--2522, Feb. 2024.
\newblock \doi{10.1093/mnras/stae169}.

\bibitem[{Thomson} et~al.(2023){Thomson}, {McConnell}, {Lenc}, {Galvin}, {Rudnick}, {Heald}, {Hale}, {Duchesne}, {Anderson}, {Carretti}, {Federrath}, {Gaensler}, {Harvey-Smith}, {Haverkorn}, {Hotan}, {Ma}, {Murphy}, {McClure-Griffiths}, {Moss}, {O'Sullivan}, {Raja}, {Seta}, {Van Eck}, {West}, {Whiting}, and {Wieringa}]{Thomson+23}
A.~J.~M. {Thomson} et al.
\newblock \emph{\pasa}, 40:\penalty0 e040, Aug. 2023.
\newblock \doi{10.1017/pasa.2023.38}.

\bibitem[{Thomson} et~al.(2026){Thomson}, {Galvin}, {Duchesne}, {Lenc}, {Heald}, {Hlinka}, {Malik}, {Anderson}, {Osinga}, {Baidoo}, {McClure-Griffiths}, {Hutschenreuter}, {O'Sullivan}, {Akahori}, {Gaensler}, {Leahy}, {Ma}, {Moss}, {Rudnick}, {Van Eck}, and {West}]{Thomson+26}
A.~J.~M. {Thomson} et al.
\newblock \emph{arXiv e-prints}, art. arXiv:2605.16917, May 2026.
\newblock \doi{10.48550/arXiv.2605.16917}.

\bibitem[{Tram} and {Hoang}(2022)]{Tram+22}
L.~N. {Tram} and T.~{Hoang}.
\newblock \emph{Frontiers in Astronomy and Space Sciences}, 9:\penalty0 923927, Oct. 2022.
\newblock \doi{10.3389/fspas.2022.923927}.

\bibitem[{Truong} and {Hoang}(2025)]{TruongHoang2025}
B.~{Truong} and T.~{Hoang}.
\newblock \emph{\apj}, 981\penalty0 (1):\penalty0 83, Mar. 2025.
\newblock \doi{10.3847/1538-4357/adb423}.

\bibitem[{Unger} and {Farrar}(2024)]{Unger+24}
M.~{Unger} and G.~R. {Farrar}.
\newblock \emph{\apj}, 970\penalty0 (1):\penalty0 95, July 2024.
\newblock \doi{10.3847/1538-4357/ad4a54}.

\bibitem[{Van Eck} et~al.(2011){Van Eck}, {Brown}, {Stil}, {Rae}, {Mao}, {Gaensler}, {Shukurov}, {Taylor}, {Haverkorn}, {Kronberg}, and {McClure-Griffiths}]{VanEck+11}
C.~L. {Van Eck} et al.
\newblock \emph{\apj}, 728\penalty0 (2):\penalty0 97, Feb. 2011.
\newblock \doi{10.1088/0004-637X/728/2/97}.

\bibitem[{Van Eck} et~al.(2023){Van Eck}, {Gaensler}, {Hutschenreuter}, {Livingston}, {Ma}, {Riseley}, {Thomson}, {Adebahr}, {Basu}, {Birkinshaw}, {En{\ss}lin}, {Heald}, {Mao}, and {McClure-Griffiths}]{VanEck+23}
C.~L. {Van Eck} et al.
\newblock \emph{\apjs}, 267\penalty0 (2):\penalty0 28, Aug. 2023.
\newblock \doi{10.3847/1538-4365/acda24}.

\bibitem[{Van Eck} et~al.(2026){Van Eck}, {R. Purcell}, {Baidoo}, {Thomson}, {Ma}, {Oberhelman}, {Osinga}, {Vanderwoude}, {West}, {Ideguchi}, {Par{\'e}}, {Kaczmarek}, {Willis}, {Akahori}, {Anderson}, {Gaensler}, {O'Sullivan}, {Sun}, {Amaral}, {Riseley}, {Stil}, and {Zhang}]{vaneck+26}
C.~L. {Van Eck} et al.
\newblock \emph{\apjs}, 283\penalty0 (1):\penalty0 28, Mar. 2026.
\newblock \doi{10.3847/1538-4365/ae3dea}.

\bibitem[{Vanderwoude} et~al.(2024){Vanderwoude}, {West}, {Gaensler}, {Rudnick}, {Van Eck}, {Thomson}, {Andernach}, {Anderson}, {Carretti}, {Heald}, {Leahy}, {McClure-Griffiths}, {O'Sullivan}, {Tahani}, and {Willis}]{Vanderwoude+24}
S.~{Vanderwoude} et al.
\newblock \emph{\aj}, 167\penalty0 (5):\penalty0 226, May 2024.
\newblock \doi{10.3847/1538-3881/ad2fc8}.

\bibitem[{Vernstrom} et~al.(2019){Vernstrom}, {Gaensler}, {Rudnick}, and {Andernach}]{Vernstrom+19}
T.~{Vernstrom}, B.~M. {Gaensler}, L.~{Rudnick}, and H.~{Andernach}.
\newblock \emph{\apj}, 878\penalty0 (2):\penalty0 92, June 2019.
\newblock \doi{10.3847/1538-4357/ab1f83}.

\bibitem[{West} et~al.(2020){West}, {Henriksen}, {Ferri{\`e}re}, {Woodfinden}, {Jaffe}, {Gaensler}, and {Irwin}]{West+20}
J.~L. {West} et al.
\newblock \emph{\mnras}, 499\penalty0 (3):\penalty0 3673--3689, Dec. 2020.
\newblock \doi{10.1093/mnras/staa3068}.

\bibitem[{West} et~al.(2021){West}, {Landecker}, {Gaensler}, {Jaffe}, and {Hill}]{West+21}
J.~L. {West} et al.
\newblock \emph{\apj}, 923\penalty0 (1):\penalty0 58, Dec. 2021.
\newblock \doi{10.3847/1538-4357/ac2ba2}.

\bibitem[{Wieringa} et~al.(1993){Wieringa}, {de Bruyn}, {Jansen}, {Brouw}, and {Katgert}]{Wieringa+93}
M.~H. {Wieringa} et al.
\newblock \emph{\aap}, 268:\penalty0 215--229, Feb. 1993.

\bibitem[{Wolleben} et~al.(2019){Wolleben}, {Landecker}, {Carretti}, {Dickey}, {Fletcher}, {McClure-Griffiths}, {McConnell}, {Thomson}, {Hill}, {Gaensler}, {Han}, {Haverkorn}, {Leahy}, {Reich}, and {Taylor}]{Wolleben+19}
M.~{Wolleben} et al.
\newblock \emph{\aj}, 158\penalty0 (1):\penalty0 44, July 2019.
\newblock \doi{10.3847/1538-3881/ab22b0}.

\bibitem[{Wolleben} et~al.(2021){Wolleben}, {Landecker}, {Douglas}, {Gray}, {Ordog}, {Dickey}, {Hill}, {Carretti}, {Brown}, {Gaensler}, {Han}, {Haverkorn}, {Kothes}, {Leahy}, {McClure-Griffiths}, {McConnell}, {Reich}, {Taylor}, {Thomson}, and {West}]{Wolleben+21}
M.~{Wolleben} et al.
\newblock \emph{\aj}, 162\penalty0 (1):\penalty0 35, July 2021.
\newblock \doi{10.3847/1538-3881/abf7c1}.

\bibitem[{Wu} et~al.(2009){Wu}, {Kim}, {Ryu}, {Cho}, and {Alexander}]{Wu+09}
Q.~{Wu} et al.
\newblock \emph{\apjl}, 705\penalty0 (1):\penalty0 L86--L89, Nov. 2009.
\newblock \doi{10.1088/0004-637X/705/1/L86}.

\bibitem[{Xu} et~al.(2022){Xu}, {Han}, {Wang}, and {Yan}]{Xu+22}
J.~{Xu}, J.~{Han}, P.~{Wang}, and Y.~{Yan}.
\newblock \emph{Science China Physics, Mechanics, and Astronomy}, 65\penalty0 (12):\penalty0 129704, Dec. 2022.
\newblock \doi{10.1007/s11433-022-2033-2}.

\bibitem[{Yao} et~al.(2017){Yao}, {Manchester}, and {Wang}]{Yao+17}
J.~M. {Yao}, R.~N. {Manchester}, and N.~{Wang}.
\newblock \emph{\apj}, 835\penalty0 (1):\penalty0 29, Jan. 2017.
\newblock \doi{10.3847/1538-4357/835/1/29}.

\end{thebibliography}
\end{document}